\documentclass[fleqn,twoside,twocolumn,nofootinbib]{revtex4}
\usepackage{graphicx}
\usepackage{amsmath}
\usepackage{amssymb}
\usepackage{amstext}

\begin{document}

\title{Exciton insulator states for particle-hole pair in ZnO/(Zn,Mg)O quantum wells and for Dirac cone.}

\author{Lyubov E. Lokot}

\affiliation{Institute of Semiconductor Physics, NAS of Ukraine, 41, Nauky Ave., Kyiv 03028, Ukraine, e-mail: llokot@gmail.com}

\begin{abstract}

In this paper a theoretical studies of the space separation of electron and hole wave functions in the quantum well $\textrm{ZnO}/\textrm{Mg}_{0.27}\textrm{Zn}_{0.73}\textrm{O}$ are presented. For this aim the self-consistent solution of the Schr\"{o}dinger equations for electrons and holes and the Poisson equations at the presence of spatially varying quantum well potential due to the piezoelectric effect and local exchange-correlation potential is found. The one-dimensional Poisson equation contains the Hartree potential which includes the one-dimensional charge density for electrons and holes along the polarization field distribution. The three-dimensional Poisson equation contains besides the one-dimensional charge density for electrons and holes the exchange-correlation potential which is built on convolutions of a plane-wave part of wave functions in addition. In ZnO/(Zn,Mg)O quantum well the electron-hole pairing leads to the exciton insulator states. An exciton insulator states with a gap 3.4 eV are predicted. If the electron and hole are separated, their energy is higher on 0.2 meV than if they are paired. The particle-hole pairing leads to the Cooper instability. In the paper a theoretical study the both the quantized energies of excitonic states and their wave functions in graphene is presented. An integral two-dimensional Schr\"{o}dinger equation of the electron-hole pairing for a particles with electron-hole symmetry of reflection is exactly solved. The solutions of Schr\"{o}dinger equation in momentum space in graphene by projection the two-dimensional space of momentum on the three-dimensional sphere are found exactly. We analytically solve an integral two-dimensional Schr\"{o}dinger equation of the electron-hole pairing for particles with electron-hole symmetry of reflection. In single-layer graphene (SLG) the electron-hole pairing leads to the exciton insulator states. Quantized spectral series and light absorption rates of the excitonic states which distribute in valence cone are found exactly. If the electron and hole are separated, their energy is higher than if they are paired. The particle-hole symmetry of Dirac equation of layered materials allows perfect pairing between electron Fermi sphere and hole Fermi sphere in the valence cone and conduction cone and hence driving the Cooper instability.

PACS number(s): 73.21.Fg, 77.22.Ej, 78.20.H-, 78.66.Hf, 81.05.ue, 81.05.U-, 71.30.+h, 71.10.-w.

\end{abstract}

\maketitle

\section{Introduction}

The zinc oxides present a new state of matter where the electron-hole pairing leads to the exciton insulator states ~\cite{Jerome}. The Coulomb interaction leads to the electron-hole bound states scrutiny study of which acquire significant attention in the explanations of high-temperature superconductivity.

There has been widely studied in the blue, ultraviolet spectral ranges lasers based on direct wide-bandgap hexagonal w\"{u}rtzite crystal material systems such as ZnO ~\cite{{Bakin},{Zippel},{Shubina},{Li},{Tabares},{Sun}}. Significant success has been obtained in growth ZnO quantum wells with (ZnMg)O barriers by scrutinized methods of growth ~\cite{{Chauveau},{Wassner}}. The carrier relaxation from (ZnMg)O barrier layers into a ZnO quantum well through time-resolved photoluminescense spectroscopy is studied in the paper ~\cite{{Chernikov}}. The time of filling of particles for the single ZnO quantum well is found to be 3 ps ~\cite{{Chernikov}}.

In the paper we present a theoretical investigation of the intricate interaction of the electron-hole plasma with a polarization-induced electric fields. The confinement of wave functions has a strong influence on the optical properties which is observed with a dependence from the intrinsic electric field which is calculated to be 0.37 MV/cm ~\cite{{Ashrafi}}, causing to the quantum-confined Stark effect (QCSE). In this paper we present the results of theoretical studies of the space separation of electron and hole wave functions by self-consistent solution of the Schr\"{o}dinger equations for electrons and holes and the Poisson equations at the presence of spatially varying quantum well potential due to the piezoelectric effect and the local exchange-correlation potential.

In addition large electron and hole effective masses, large carrier densities in quantum well ZnO are of cause for population inversions. These features are comparable to GaN based systems ~\cite{{Wang},{Lokot}}.

A variational simulation in effective-mass approximation is used for the conduction band dispersion and for quantization of holes a Schr\"{o}dinger equation is solved with w\"{u}rtzite hexagonal effective Hamiltonian ~\cite{{Bir}} including deformation potentials ~\cite{{Langer}}. Keeping in mind the above mentioned equations and the potential energies which have been included in this problem from Poisson's equations we have obtained completely self-consistent band structures and wave functions.

We consider the pairing between oppositely charged particles with complex dispersion. If the exciton Bohr radius is grater than the localization range particle-hole pair, the excitons may be spontaneously created.

If the Hartree-Fock band gap energy is greater than the exciton energy in ZnO/(Zn,Mg)O quantum wells then excitons may be spontaneously created. It is known in narrow-gap semiconductor or semimetal at sufficiently low temperature the insulator ground state is instable with respect to the exciton formation ~\cite{{Stroucken},{Jerome}}, leading to a spontaneously creating of excitons ~\cite{Parfitt}. In a system undergo a phase transition into a exciton insulator phase similarly to Bardeen-Cooper-Schrieffer (BCS) superconductor.

An exciton insulator states with a gap 3.4 eV are predicted. The particle-hole pairing leads to the Cooper instability.

\section{Theoretical study}

\subsection{Effective Hamiltonian}

\begin{widetext}
It is known
\cite{{Bir},{Yu}} that the valence-band
spectrum of hexagonal w\"{u}rtzite crystal at the $\Gamma$ point originates from the sixfold
degenerate $\Gamma_{15}$ state.  Under the action of the hexagonal
crystal field in w\"{u}rtzite
crystals, $\Gamma_{15}$ splits and leads to the formation of two
levels: $\Gamma_{1}$, $\Gamma_{5}$. The wave functions of the valence band transform according to
the representation $\Gamma_{1}+\Gamma_{5}$ of the point group $C_{6v}$, while the wave function of the conduction band
transforms according to the representation $\Gamma_{1}$.

\begin{tabular}{cccccccc} \hline\hline
\multicolumn{1}{c}{$C_{6v}$} &
\multicolumn{1}{c}{$E$} &
\multicolumn{1}{c}{$C_{2}$} &
\multicolumn{1}{c}{$2C_{3}$} &
\multicolumn{1}{c}{$2C_{6}$} &
\multicolumn{1}{c}{$3\sigma_{v}$} &
\multicolumn{1}{c}{$3\sigma_{v}'$} &
\multicolumn{1}{c}{$ $} \\
\hline
$\Gamma_{1}+\Gamma_{5}$ & 3 & -1 & 0 & 2 & 1 & 1 & $ $ \\
$g^{2}$ & $E$ & $E$ & $C_{3}$ & $C_{3}$ & $E$ & $E$ & $ $ \\
$\chi^{2}_{\psi}(g)$ & 9 & 1 & 0 & 4 & 1 & 1 & $ $\\
$\chi_{\psi}(g^{2})$ & 3 & 3 & 0 & 0 & 3 & 3 & $ $\\
$\frac{1}{2}[\chi_{\psi}^{2}(g)+\chi_{\psi}(g^{2})]$ & 6 & 2 & 0 & 2 & 2 & 2 & $2\Gamma_{1}+\Gamma_{5}+\Gamma_{6}$\\
$\frac{1}{2}\{\chi_{\psi}^{2}(g)-\chi_{\psi}(g^{2})\}$ & 3 & -1 & 0 & 2 & -1 & -1 & $\Gamma_{2}+\Gamma_{5}$\\ \hline\hline
\end{tabular}
\end{widetext}

An irreducible presentations for orbital angular momentum $j$ may be built from formula

\begin{equation}
\chi_{j}(\varphi)=\frac{\sin{(j+\frac{1}{2})\varphi}}{\sin{\frac{\varphi}{2}}}.
\end{equation}
For the vector representational $j=1$
\begin{equation}
\chi_{v}(\varphi)=\frac{\sin{\frac{3\varphi}{2}}}{\sin{\frac{\varphi}{2}}}=1+2\cos{\varphi}.
\end{equation}

\begin{widetext}
\begin{tabular}{cccccccc} \hline\hline
\multicolumn{1}{c}{$C_{6v}$} &
\multicolumn{1}{c}{$E$} &
\multicolumn{1}{c}{$C_{2}$} &
\multicolumn{1}{c}{$2C_{3}$} &
\multicolumn{1}{c}{$2C_{6}$} &
\multicolumn{1}{c}{$3\sigma_{v}$} &
\multicolumn{1}{c}{$3\sigma_{v}'$} &
\multicolumn{1}{c}{$ $} \\
\hline
\multicolumn{1}{c}{$\chi_{v}$} &
\multicolumn{1}{c}{$3$} &
\multicolumn{1}{c}{$-1$} &
\multicolumn{1}{c}{$0$} &
\multicolumn{1}{c}{$2$} &
\multicolumn{1}{c}{$1$} &
\multicolumn{1}{c}{$1$} &
\multicolumn{1}{c}{$\Gamma_{1}+\Gamma_{5}$} \\
$\frac{1}{2}[\chi_{v}^{2}(g)+\chi_{v}(g^{2})]$ & 6 & 2 & 0 & 2 & 2 & 2 & $2\Gamma_{1}+\Gamma_{5}+\Gamma_{6}$\\
$\frac{1}{2}\{\chi_{v}^{2}(g)-\chi_{v}(g^{2})\}$ & 3 & -1 & 0 & 2 & -1 & -1 & $\Gamma_{2}+\Gamma_{5}$\\ \hline\hline
\end{tabular}
\end{widetext}

The direct production of two irreducible presentations of wave function and wave vector of difference $\kappa-\Gamma$ expansion with taken into account time inversion can be expanded on

\begin{equation}
\begin{array}{c}
p^{\alpha}: \tau_{v}\times\,\tau_{\psi}=(\Gamma_{1}+\Gamma_{5})\times\,(\Gamma_{2}+\Gamma_{5})=\\
=\Gamma_{5}\times\Gamma_{5},
\end{array}
\end{equation}
for the square of wave vector

\begin{equation}
\begin{array}{c}
[p^{\alpha}p^{\beta}]: \tau_{v^{2}}\times\,\tau_{\psi}=(2\Gamma_{1}+\Gamma_{5}+\Gamma_{6})\times\,(2\Gamma_{1}+\Gamma_{5}+\Gamma_{6})=\\
=4\Gamma_{1}\times\Gamma_{1}+\Gamma_{5}\times\Gamma_{5}+\Gamma_{6}\times\Gamma_{6}.
\end{array}
\end{equation}

In the low-energy limit the Hamiltonian of w\"{u}rtzite

\begin{equation}
\begin{array}{c}
\hat{H}_{0}=I(\Delta_{1}+\Delta_{2})+\\
+\Delta_{1}J_{z}^{2}+\Delta_{2}J_{z}\sigma_{z}+\sqrt{2}\Delta_{3}(J_{+}\sigma_{-}+J_{-}\sigma_{+}),
\end{array}
\end{equation}

\begin{equation}
\begin{array}{c}
\hat{H}_{k}=A_{1}k_{z}^{2}+A_{2}k_{t}^{2}+(A_{3}k_{z}^{2}+A_{4}k_{t}^{2})J_{z}^{2}+\\
+A_{5}k_{z}(2[J_{z}J_{+}]k_{-}+2[J_{z}J_{-}]k_{+})+\\
+A_{6}(J_{+}^{2}k_{-}^{2}+J_{-}^{2}k_{+}^{2})+iA_{7}(J_{+}k_{-}-J_{-}k_{+}),
\end{array}
\end{equation}

\begin{equation}
\begin{array}{c}
\hat{H}_{\varepsilon}=D_{1}\varepsilon_{zz}+D_{2}\varepsilon_{\bot}^{2}+(D_{3}\varepsilon_{zz}+D_{4}\varepsilon_{\bot})J_{z}^{2}+\\
+D_{5}(2[J_{z}J_{+}]\varepsilon_{-z}+2[J_{z}J_{-}]\varepsilon_{+z})+\\
+D_{6}(J_{+}^{2}\varepsilon_{-}+J_{-}^{2}\varepsilon_{+}).
\end{array}
\end{equation}

In the basis of spherical wave functions with the
orbital angular momentum $l=1$ and the eigenvalue $m_{l}$ of its $z$
component:

\begin{widetext}
\begin{equation}
\begin{array}{c}
|1,\varsigma_{v}\rangle=\frac{1}{\sqrt{2}}(Y_{1}^{1}\psi(1/2)e^{-3i\varphi/2}e^{-3i\pi/4}\pm\,Y_{1}^{-1}\psi(-1/2)e^{3i\varphi/2}e^{3i\pi/4})\\
|2,\varsigma_{v}\rangle=\frac{1}{\sqrt{2}}(\pm\,Y_{1}^{1}\psi(-1/2)e^{-i\varphi/2}e^{-i\pi/4}+\,Y_{1}^{-1}\psi(1/2)e^{i\varphi/2}e^{i\pi/4})\\
|3,\varsigma_{v}\rangle=\frac{1}{\sqrt{2}}(\pm\,Y_{1}^{0}\psi(1/2)e^{-i\varphi/2}e^{-i\pi/4}+\,Y_{1}^{0}\psi(-1/2)e^{i\varphi/2}e^{i\pi/4})\\
\end{array},
\end{equation}
\end{widetext}
the Hamiltonian may be transformed to the diagonal form indicating two spin degeneracy ~\cite{Chuang}:

\begin{equation}
H_{\pm}=\left\|
\begin{array}{cccc}
F & K_{t} & \mp\,iH_{t}\\
K_{t} & G & \Delta\mp\,iH_{t}\\
\pm\,iH_{t} & \Delta\pm\,iH_{t} & \lambda\\
\end{array}
\right\|\begin{array}{cccc}
|1,\varsigma_{v}\rangle\\
|2,\varsigma_{v}\rangle\\
|3,\varsigma_{v}\rangle\\
\end{array},
\end{equation}
where
$F=\Delta_{1}+\Delta_{2}+\lambda+\theta$, $G=\Delta_{1}-\Delta_{2}+\lambda+\theta$, $\lambda=\lambda_{k}+\lambda_{\epsilon}$, $\theta=\theta_{k}+\theta_{\epsilon}$, $\lambda_{k}=\frac{\hbar^{2}}{2m_{0}}(A_{1}k_{z}^{2}+A_{2}k_{t}^{2})$, $\lambda_{\epsilon}=D_{1}\epsilon_{zz}+D_{2}(\epsilon_{xx}+\epsilon_{yy})$, $\theta_{k}=\frac{\hbar^{2}}{2m_{0}}(A_{3}k_{z}^{2}+A_{4}k_{t}^{2})$, $\theta_{\epsilon}=D_{3}\epsilon_{zz}+D_{4}(\epsilon_{xx}+\epsilon_{yy})$, $K_{t}=\frac{\hbar^{2}}{2m_{0}}(A_{5}k_{t}^{2})$, $H_{t}=\frac{\hbar^{2}}{2m_{0}}(A_{6}k_{t}k_{z})$, $\Delta=\sqrt{2}\Delta_{3}$, $k_{t}^{2}=k_{x}^{2}+k_{y}^{2}$.

From Kane model one can define the band-edge parameters such as the crystal-field splitting energy $\Delta_{cr}$, the spin-orbit splitting energy $\Delta_{so}$ and the  momentum-matrix elements for the longitudinal ($\textbf{e}\parallel{z}$) z-polarization and the transverse ($\textbf{e}\perp{z}$) polarization : $P_{z}\equiv\langle\,S|\hat{p}_{z}|Z\rangle$, $P_{\perp}\equiv\langle\,S|\hat{p}_{x}|X\rangle\equiv\langle\,S|\hat{p}_{y}|Y\rangle$. Here we use the effective-mass parameters, energy splitting parameters, deformation potential parameters as in papers ~\cite{{Langer},{Yan},{Madelung}}.

We consider a quantum well of width $w$ in ZnO under biaxial strain, which
is oriented perpendicularly to the growth direction (0001) and
localized in the spatial region $-w/2<z<w/2$. In the ZnO/MgZnO quantum
well structure, there is a strain-induced electric field. This
piezoelectric field, which is perpendicular to the quantum well
plane (i.e., in z direction) may be appreciable because of the large
piezoelectric constants in w\"{u}rtzite structures.

The transverse components of the biaxial strain are proportional to the difference between the lattice constants of materials of the well and the barrier and depend on the Mg content x: $\epsilon_{xx}=\epsilon_{yy}=\frac{a_{Mg_{x}Zn_{1-x}O}-a_{ZnO}}{a_{ZnO}}$, $a_{Mg_{x}Zn_{1-x}O}=a_{ZnO}+x\,(a_{MgO}-a_{ZnO})$, $a_{ZnO}=0.32496$ nm, $a_{MgO}=0.4216$ nm ~\cite{{Madelung}}. The longitudinal component of a deformation is expressed through elastic constants and the transverse component of a deformation: $\epsilon_{zz}=-2\frac{C_{13}}{C_{33}}\epsilon_{xx}$.

The physical parameters for ZnO are as follows. We take the effective-mass parameters ~\cite{{Yan}}: $A_{1}=-2.743$, $A_{2}=-0.393$, $A_{3}=2.377$, $A_{4}=-2.069$, $A_{5}=-2.051$, $A_{6}=-2.099$, $m_{e}^{z,\perp}=0.329 m_{0}$, where $m_{0}$ is the
electron rest mass in the vacuum, the parameters for deformation potential ~\cite{{Langer}}: $D_{1}=-3800$ meV, $D_{2}=-3800$ meV, $D_{3}=-800$ meV, $D_{4}=1400$ meV, $D_{cz} := -6860$ meV, $D_{c\bot} := -6260$ meV, and the energy parameters at 300 K ~\cite{{Yan},{Madelung}}: $E_{g}=3400$ meV, $\Delta_{1}=\Delta_{cr}=36.3$ meV, $\Delta_{2}/3=0.63$ meV, $\Delta_{3}/3=2.47$ meV, $\Delta_{2}=\Delta_{3}=\Delta_{so}$ the elastic constant ~\cite{{Madelung}}: $C_{13}=90$ GPa and $C_{33}=196$ GPa, the permittivity of the host materials $\kappa=7.8$.

\subsection{ZnO/(Zn,Mg)O quantum well}

We take the following wave functions written as vectors in the three-dimensional Bloch space:
\begin{equation}
|\nu\,\varsigma_{v}\,k_{t}\rangle=\left\|
\begin{array}{cccc}
\sum_{i=1}^{m}\Psi_{k_{t}}^{(1)}[i,\nu]\,\psi_{i}(Z)\\
\sum_{i=1}^{m}\Psi_{k_{t}}^{(2)}[i,\nu]\,\psi_{i}(Z)\\
\sum_{i=1}^{m}\Psi_{k_{t}}^{(3)}[i,\nu]\,\psi_{i}(Z)\\
\end{array}
\right\|\begin{array}{cccc}
|1,\varsigma_{v}\rangle\\
|2,\varsigma_{v}\rangle\\
|3,\varsigma_{v}\rangle\\
\end{array}.\end{equation}
The Bloch vector of $\nu$-type hole with spin $\varsigma_{v}=\pm\,1/2$ and momentum $k_{t}$ is specified by its three coordinates $[\Psi_{k_{t}}^{(1)}[m,\nu],\,\Psi_{k_{t}}^{(2)}[m,\nu],\,\Psi_{k_{t}}^{(3)}[m,\nu]]$ in the basis $[|1,\varsigma_{v}\rangle,\,|2,\varsigma_{v}\rangle,\,|3,\varsigma_{v}\rangle]$ ~\cite{{Chuang}}, known as spherical harmonics with the orbital angular momentum $l=1$ and the eigenvalue $m_{l}$ its $z$ component. The envelope $Z$-dependent part of the quantum well eigenfunctions can be specified from the boundary conditions $\psi_{m}(Z=0)=\psi_{m}(Z=1)=0$ of the infinite quantum well as
\begin{equation}
\psi_{m}(Z)=\sqrt{\frac{2}{w}}\,\sin{(\pi\,m\,Z)},
\end{equation}
where $Z=(\frac{z}{w}+\frac{1}{2})$, $m$ is a natural number.
Thus the hole wave function can be written as
\begin{equation}
\Psi_{\nu\,\varsigma_{v}\,k_{t}}(\textbf{r})=\frac{e^{i\,k_{t}\,\rho_{t}}}{\sqrt{A}}\,|\nu\,\varsigma_{v}\,k_{t}\rangle.
\end{equation}

The valence subband structure $E_{\nu}^{\varsigma_{v}}(k_{t})$ can be determined by solving equations system:
\begin{equation}
\begin{array}{c}
\sum_{j=1}^{3}(H_{ij}^{\varsigma_{v}}(k_{z}=-i\,\frac{\partial}{\partial{z}})+V(z)+\delta_{ij}E_{\nu}^{\varsigma_{v}}(k_{t}))\times\\
\times\phi_{\nu}^{(j)\varsigma_{v}}(z,k_{t})=0,\\
\end{array}
\end{equation}
where $\phi_{\nu}^{(j)\varsigma_{v}}(z,k_{t})=\sum_{n=1}^{m}\Psi_{k_{t}}^{(j)}[n,\nu]\,\psi_{n}(z)$, $i=1,2,3$.

The wave function of electron of first energy level with accounts QCSE ~\cite{{Bastard}}:
\begin{equation}
\Psi(\textbf{r})=\frac{1}{\sqrt{A}}e^{ik_{t}\rho}\Psi(Z,\xi)|S\rangle|\varsigma_{c}\rangle,
\end{equation}
where
\begin{widetext}
\begin{equation}
\begin{array}{c}
\Psi(Z,\xi)=\begin{cases}
\psi_{1}(Z,\xi)=C_{1}e^{(\kappa_{0}-\xi)(w\,Z)}, Z\in(-\infty..0)\cr
\psi(Z,\xi)=C\sin{(k_{0}w\,(Z-\frac{1}{2})+\delta_{0})}e^{-\xi\,w\,(Z-\frac{1}{2})},Z\in[0..1]\cr
\psi_{2}(Z,\xi)=C_{2}e^{-(\kappa_{0}+\xi)w(Z-1)}, Z\in(1..\infty).
\end{cases}
\end{array},
\end{equation}
\end{widetext}
$|S\rangle=Y_{0}^{0}$, $\varsigma_{c}=\pm\,1/2$.

\begin{figure}
\includegraphics*[bb=5 10 1000 600,width=4in]{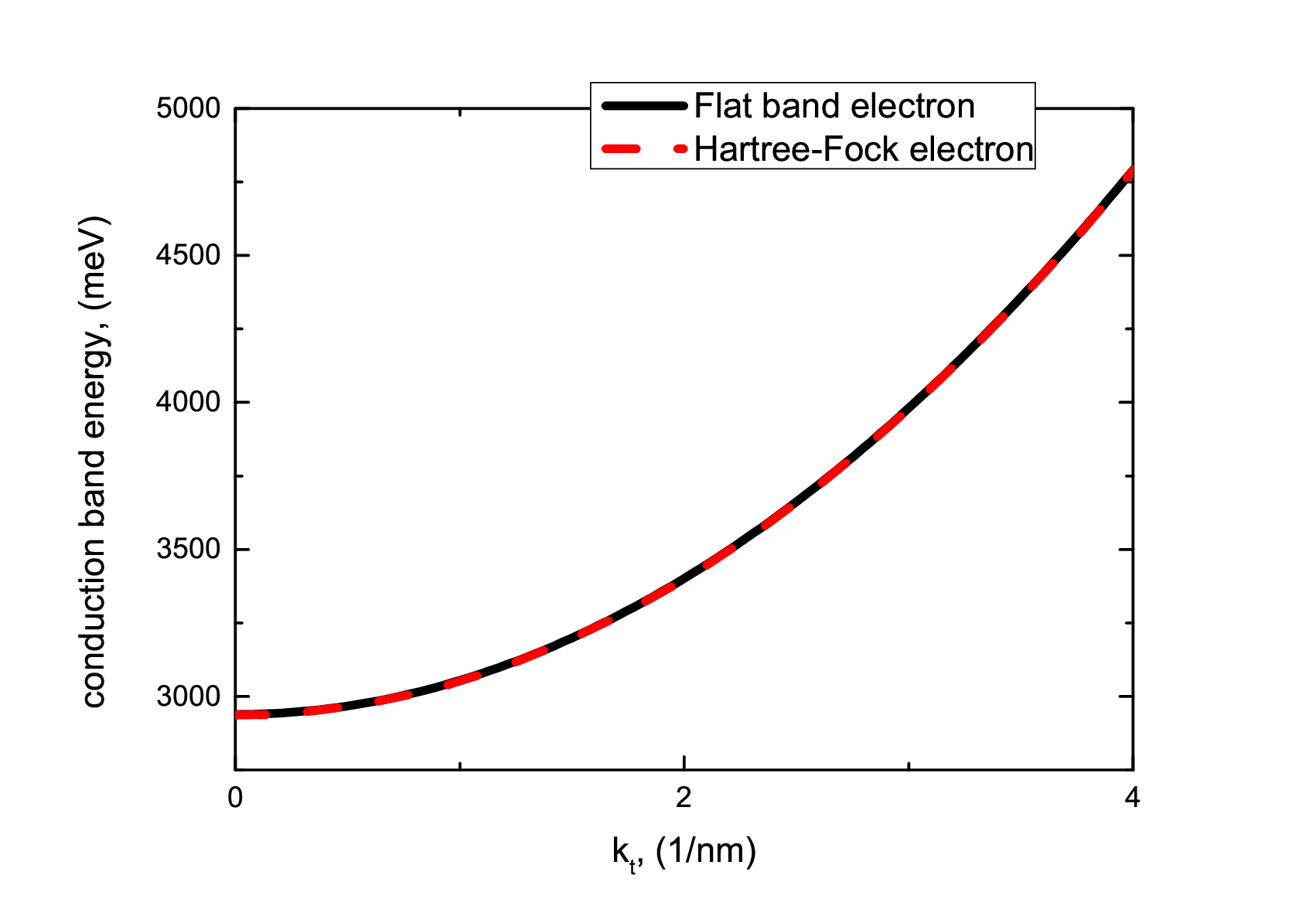}
\includegraphics*[bb=5 10 1000 600,width=4in]{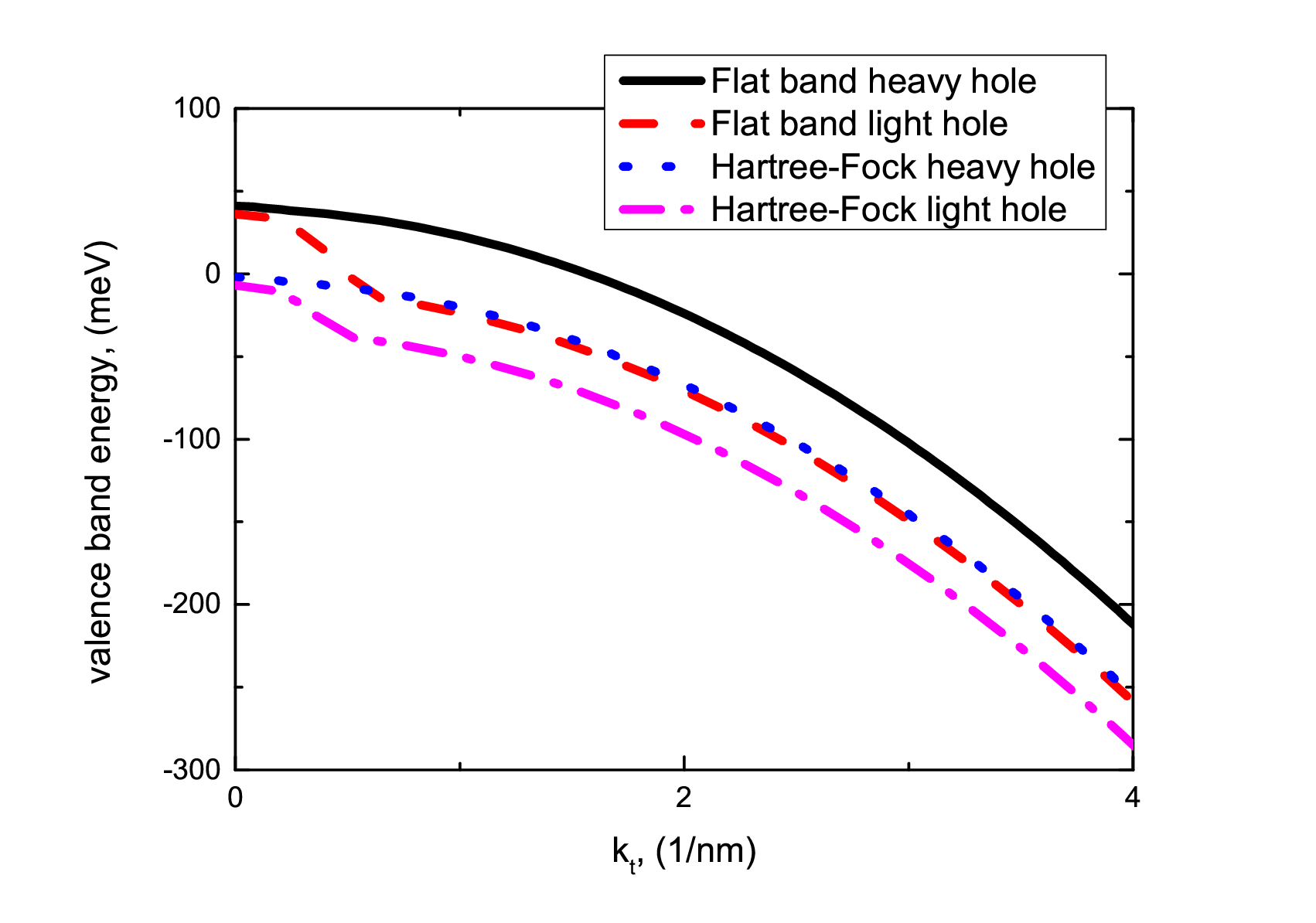}
\caption{(Color online) For the quantum well $\textrm{ZnO}/\textrm{Mg}_{0.27}\textrm{Zn}_{0.73}\textrm{O}$ with a width 6 nm, at a carriers concentration $4*10^{12}$ $\textrm{cm}^{-2}$, (a) conduction band energy; (b) valence band energy.}
\end{figure}

\begin{figure}
\includegraphics*[bb=5 10 1000 600,width=4in]{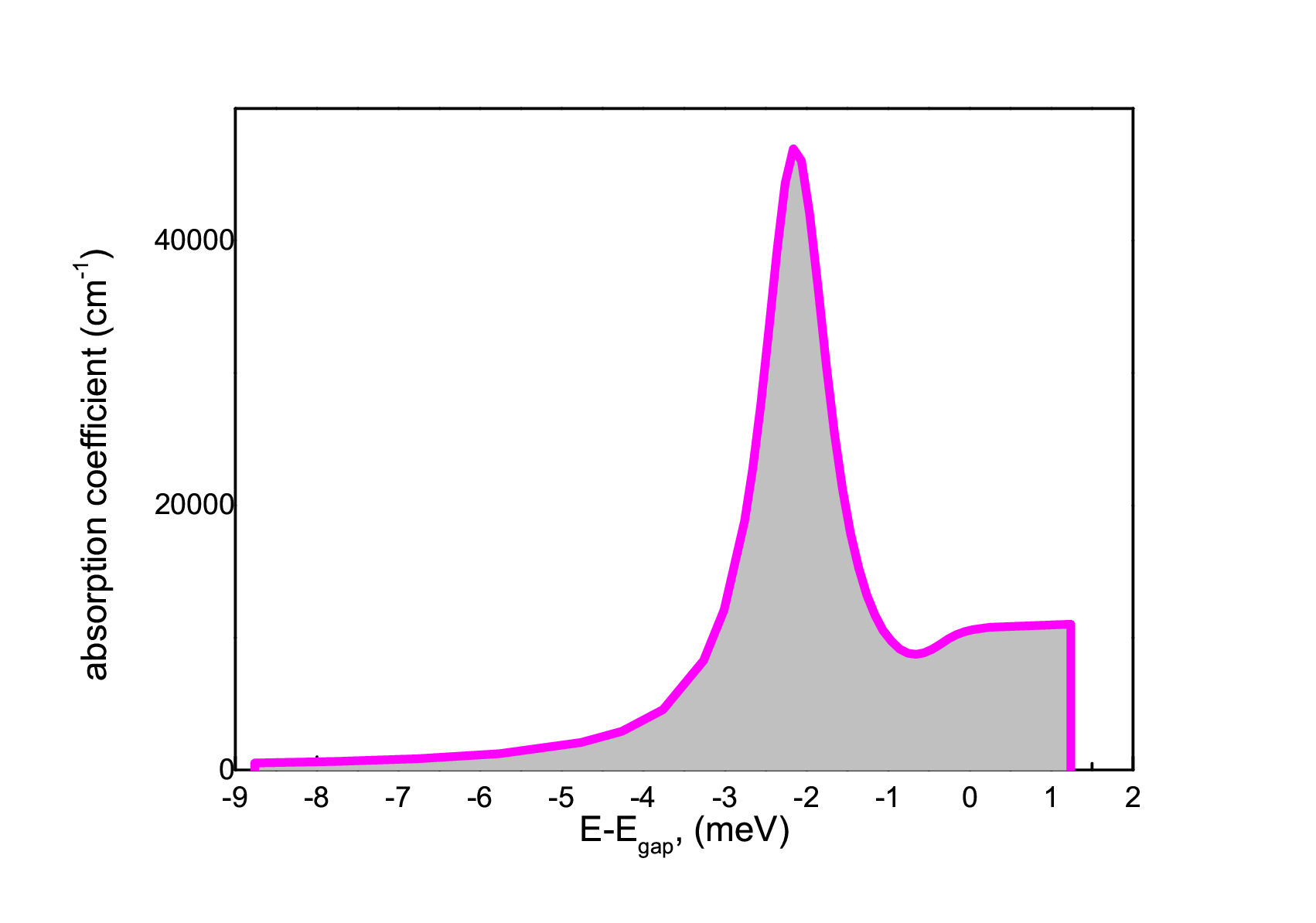}
\caption{(Color online) Absorption coefficient for the quantum well $\textrm{ZnO}/\textrm{Mg}_{0.27}\textrm{Zn}_{0.73}\textrm{O}$ with a width 6 nm, at a carriers concentration $4*10^{12}$ $\textrm{cm}^{-2}$, at a temperature 310 K.}
\end{figure}

From bond conditions ~\cite{{Bastard},{Landau}} $\psi_{1}(Z,\xi)|_{Z=0}=\psi(Z,\xi)|_{Z=0}$, $\psi_{2}(Z,\xi)|_{Z=1}=\psi(Z,\xi)|_{Z=1}$, $\frac{\psi'_{1}(Z,\xi)}{\psi_{1}(Z,\xi)}|_{Z=0}=\frac{\psi'(Z,\xi)}{\psi(Z,\xi)}|_{Z=0}$, $\frac{\psi'_{2}(Z,\xi)}{\psi_{2}(Z,\xi)}|_{Z=1}=\frac{\psi'(Z,\xi)}{\psi(Z,\xi)}|_{Z=1}$, one can find
$C_{1}=C\sin{(-\frac{k_{0}w}{2}+\delta_{0})}e^{\xi\frac{w}{2}}$, $C_{2}=C\sin{(\frac{k_{0}w}{2}+\delta_{0})}e^{-\xi\frac{w}{2}}$,
$\kappa_{0}=k_{0}(\frac{1-\cos{k_{0}w}}{\sin{k_{0}w}})$, $\delta_{0}=\frac{k_{0}w}{2}+\arctan{\frac{\kappa_{0}}{k_{0}}}$, where A is the area of the quantum well in the $xy$ plane, $\rho$ is the two-dimensional vector in the $xy$ plane, $k_{t}=(k_{x},k_{y})$ is in-plane wave vector. The constant multiplier $C$ is found from normalization condition:
\begin{equation}
\int_{-\infty}^{\infty}|\Psi(Z,\xi)|^{2}wdZ=1.
\end{equation}

One can find the functional, which is built in the form:
\begin{equation}
J(\xi)=\frac{\langle\Psi|\hat{H}|\Psi\rangle}{\langle\Psi|\Psi\rangle},
\end{equation}
where
\begin{equation}
H=H_{c}+V(z),
\end{equation}
where $H_{c}$ is a conduction band kinetic energy including deformation potential:
\begin{equation}
\begin{array}{c}
H_{c}=E_{g}+\Delta_{1}+\Delta_{2}+\frac{\hbar^{2}}{2m_{e}^{\bot}}k_{t}^{2}-\frac{\hbar^{2}}{2m_{e}^{z}}\frac{\partial^{2}}{\partial\,z^{2}}+\\
+D_{cz}\epsilon_{zz}+D_{c\bot}(\epsilon_{xx}+\epsilon_{yy}).
\end{array}
\end{equation}

The potential energies $V(z)$ can look for as follows:
\begin{equation}
V(z)=e\Phi^{H}(z)+\delta\,U_{c,v}(z)+\Phi_{xc}(z),
\end{equation}
where $\Phi^{H}(z)$ is the solution of one-dimensional Poisson's equation with the strain-induced electric field in the quantum well, $\delta\,U_{c,v}(z)$ are the conduction and valence bandedge discontinuities which can be represented in the form ~\cite{{Makino}}:
\begin{equation}
\delta\,U_{c}(z)=
\begin{cases}
U_{0}-eEw(\frac{z}{w}+1), z\in(-\infty..-w/2)\cr
eEz, z\in[-w/2..w/2]\cr
U_{0}-eEw(\frac{z}{w}-1), z\in(w/2..\infty).
\end{cases}
\end{equation}
$\Phi_{xc}(z)$ is exchange-correlation potential energy which is found from the solution of three-dimensional Poisson's equation, using both an expression by Gunnarsson and Lundquist ~\cite{{Gunnarsson}}, and following criterions. At carrier densities $4*10^{12}$ $\textrm{cm}^{-2}$, the criterion $k_{F}>\sqrt{n}/4$ at a temperature T=0 K as $1>0.1$ has been carried. $k_{F}$ is Fermi wave vector. The criterion does not depend from a width of well. The ratio of Coulomb potential energy to the Fermi energy is $r_{s}=E_{C}/E_{F}=0.63<1$.
The problem consists of the one-dimensional Poisson's equation solving of which may be found Hartree potential energy and three-dimensional Poisson's equation which is separated on one-dimensional and two-dimensional equations by separated of variables using a criterion $[\Psi_{\alpha,\nu,n}(k_{F},z)\,\sin{\textbf{k}_{F}\boldsymbol
{\rho}}]<<1$, where $\alpha={e,h}$. The three-dimensional Poisson's equation includes local exchange-correlation potential:
\begin{equation}
\frac{d^{2}\Phi_{e,h}}{dz^{2}}+\Delta_{\rho}\Phi_{e,h}=\frac{4\pi}{\kappa}(\rho_{e,h}^{H}(z;g)+\rho_{e,h}^{xc}(\textbf{r},\textbf{r}')),
\end{equation}
\begin{equation}
\frac{d^{2}\Phi_{e,h}^{H}}{dz^{2}}=\frac{4\pi}{\kappa}\rho_{e,h}^{H}(z;g),
\end{equation}
\begin{equation}
\Delta_{\rho}\Phi_{e,h}^{xc}=\frac{4\pi}{\kappa}\rho_{e,h}^{xc}(\textbf{r},\textbf{r}'),
\end{equation}
where
\begin{equation}
\rho_{e,h}^{H}(z;g)=\mp\,e\sum_{\nu,n,k_{t}}|\Psi_{e,h,\nu,n}(k_{t},z)|^{2}f_{n,\nu}(k_{t};g),
\end{equation}

\begin{equation}
\begin{array}{c}
f_{n,\nu}(k_{t};g)=\frac{1}{e^{(\epsilon_{n,\nu,k_{t}}+\frac{g}{2}\sum_{i\neq\,j}\frac{1}{|\textbf{r}_{i}-\textbf{r}_{j}|}-\mu)/kT}+1}=\\
=\frac{1}{(e^{1}(1+r_{s}+r_{s}^{2}+...))^{(\epsilon_{n,\nu,k_{t}}-\mu)/kT}+1}.
\end{array}
\end{equation}

The solution of equations system (13), (17), (22) as well as (13), (17), (23) does not depend from a temperature.

Solving one-dimensional Poisson's equation (23) one can find screening polarization field and Hartree potential energy by substituting her in the Schr\"{o}dinger equations. From Schr\"{o}dinger equations wave functions and bandstructure are found. The conclusive determination of screening polarization field is determined by iterating Eqs. (13), (17), (22) until the solutions of conduction and valence band energies and wave functions are converged:
\begin{equation}
\Phi^{H}(z)=\Phi_{h}^{H}(z)+\Phi_{e}^{H}(z),
\end{equation}
\begin{widetext}
\begin{equation}
\begin{array}{c}
e\Phi_{h}^{H}(z)=\frac{2e^{2}}{\kappa}\sum_{\nu,m,l,i}g_{\nu}\int\,k_{t}dk_{t}\langle\,v_{i},\varsigma_{v}|\Psi_{k_{t}}^{i}[\nu,m]\Psi_{k_{t}}^{i}[\nu,l]|\varsigma_{v},v_{i}\rangle\,f_{\nu,p}(k_{t})\times\\
\times\begin{cases}
w(\frac{\cos{\pi\,(\frac{z}{w}+\frac{1}{2})(l+m)}}{\pi^{2}(l+m)^{2}}-\frac{\cos{\pi\,(\frac{z}{w}+\frac{1}{2})(m-l)}}{\pi^{2}(m-l)^{2}}), m\neq\,l\cr
w(\frac{(\frac{z}{w}+\frac{1}{2})^{2}}{2}+\frac{1}{4}\frac{\cos{2\pi\,m(\frac{z}{w}+\frac{1}{2})}}{\pi^{2}m^{2}}), m=l,
\end{cases}\\
\end{array}
\end{equation}
\begin{equation}
\begin{array}{c}
e\Phi_{e}^{H}(z)=-\frac{2e^{2}}{\kappa}g_{1}\int\,k_{t}dk_{t}C^{2}f_{1n}(k_{t})\times\\
\times\begin{cases}
\frac{1-\cos{(-k_{0}w+2\delta_{0})}}{2}e^{\xi\,w}\frac{e^{2(\kappa_{0}-\xi)(z+\frac{w}{2})}}{4(\kappa_{0}-\xi)^{2}}, z\in(-\infty..-w/2)\cr
\frac{e^{-2\xi\,z}}{8\xi^{2}}-\frac{2\cos{2(k_{0}z+\delta_{0})}e^{-2\xi\,z}}{(4\xi^{2}+4k_{0}^{2})^{2}}(\xi^{2}-k_{0}^{2})+\frac{\sin{2(k_{0}z+\delta_{0})}e^{-2\xi\,z}}{4(\xi^{2}+k_{0}^{2})^{2}}k_{0}\xi, z\in[-w/2..w/2]\cr
\frac{1-\cos{(k_{0}w+2\delta_{0})}}{2}e^{-\xi\,w}\frac{e^{-2(\kappa_{0}+\xi)(z-\frac{w}{2})}}{4(\kappa_{0}+\xi)^{2}}, z\in(w/2..\infty),
\end{cases}\\
\end{array}
\end{equation}
\end{widetext}
where $Z=\frac{z}{w}+\frac{1}{2}$, $g_{\nu}$ and $g_{1}$ correspond to the
degeneration  of the $\nu$ hole band and
the first quantized conduction band, respectively, $e$ is the value of
electron charge, $\kappa$ is the permittivity of a host material, and
$f_{\nu,p}(k_{t})$, $f_{1n}(k_{t})$ are the Fermi-Dirac
distributions for holes and electrons.

Exchange-correlation charge density may be determined as:
\begin{equation}
\begin{array}{c}
\rho_{e,h}^{xc}(\textbf{r},\textbf{r}')=\\
=\sum_{l=0}^{\infty}\sum_{m=-l}^{l}|\Psi_{\alpha,\nu,n}(k_{t},z)|^{2}\rho_{lm}(\boldsymbol
{\rho}-\boldsymbol
{\rho'})Y_{lm}(\frac{\boldsymbol
{\rho}-\boldsymbol
{\rho'}}{|\boldsymbol
{\rho}-\boldsymbol
{\rho'}|}),
\end{array}
\end{equation}
using the expansion of plane wave
\begin{equation}
\begin{array}{c}
\sum_{l=0}^{\infty}\sum_{m=-l}^{l}\rho_{lm}(\boldsymbol
{\rho})Y_{lm}(\frac{\boldsymbol
{\rho}}{|\boldsymbol
{\rho}|})=\\
=e^{i\,\textbf{k}_{t}\,\boldsymbol
{\rho}}=4\pi\,\sum_{l=0}^{\infty}\sum_{m=-l}^{l}i^{l}j_{l}(\textbf{k}_{t}\boldsymbol
{\rho})Y_{lm}^{*}(\frac{\textbf{k}_{t}}{k_{t}})Y_{lm}(\frac{\boldsymbol
{\rho}}{|\boldsymbol
{\rho}|}).
\end{array}
\end{equation}

At the condition $[\Psi_{\alpha,\nu,n}(k_{F},z)\,\sin{\textbf{k}_{F}\boldsymbol
{\rho}}]<<1$, the solution Eq. (24) may be found as follows
\begin{equation}
\Phi_{e,h}(xc)=\int_{0}^{\infty}\rho\rho_{00}(\boldsymbol
{\rho})\frac{1}{\rho}d\,\rho.
\end{equation}
The solution the three-dimensional Poisson's equation may be presented in the form:
\begin{equation}
\Phi_{e,h}^{xc}(z)=\Phi_{e,h}^{H}(z)\,\Phi_{e,h}(xc).
\end{equation}
The complete potential which describes piezoelectric effects and local exchange-correlation potential in quantum well one can find as follows
\begin{equation}
\Phi(z)=\Phi_{h}^{H}(z)+\Phi_{e}^{H}(z)+\Phi_{h}^{H}(z)\,\Phi_{h}(xc)+\Phi_{e}^{H}(z)\,\Phi_{e}(xc).
\end{equation}

\subsection{Uncertainty Heisenberg principle}

The excitons in semiconductors have been studied by \cite{{Vasko}}.

The Heisenberg equation for a microscopic dipole $\hat{p}_{\textbf{p}}^{\nu_{e}\nu_{h}}=\langle\,\hat{b}_{-\textbf{p}}\hat{a}_{\textbf{p}}\rangle$
due to an electron-hole pair with the electron
(hole) momentum \textbf{p} (--\textbf{p}) and the subband number
$\nu_{e}$ ($\nu_{h}$) is written in the form:

\begin{equation}
\frac{\partial\,\hat{p}_{\textbf{p}}^{\nu_{e}\nu_{h}}}{\partial\,t}=\frac{i}{\hbar}[\hat{H},\hat{p}_{\textbf{p}}^{\nu_{e}\nu_{h}}].
\end{equation}

We assume a nondegenerate situation described by the Hamiltonian $\hat{H}=\hat{H}_{0}+\hat{V}+\hat{H}_{\rm int}$, which is composed
of the kinetic energy of an electron $\epsilon_{e,\textbf{p}}^{\nu_{e}}$ and the kinetic energy
of a hole $\epsilon_{h,\textbf{p}}^{\nu_{h}}$ in the electron-hole representation:

\begin{equation}
\hat{H}_{0}=\sum_{\textbf{p}}{\epsilon_{e,\textbf{p}}^{\nu_{e}}\hat{a}_{\textbf{p}}^{+}\hat{a}_{\textbf{p}}+\epsilon_{h,\textbf{p}}^{\nu_{h}}\hat{b}_{-\textbf{p}}^{+}\hat{b}_{-\textbf{p}}},
\end{equation}
where \textbf{p} is the transversal quasimomentum of carriers in the plane of the quantum well, $\hat{a}_{\textbf{p}}$,
$\hat{a}_{\textbf{p}}^{+}$, $\hat{b}_{-\textbf{p}}$, and $\hat{b}_{-\textbf{p}}^{+}$ are the annihilation and creation
operators of an electron and a hole. The Coulomb interaction Hamiltonian for particles in the electron-hole representation takes
the form:

\[
\hat{V}=\frac{1}{2}\sum_{\textbf{p},\textbf{k},\textbf{q}}V_{q}^{\nu_{e}\nu_{e}\nu_{e}\nu_{e}}\hat{a}_{\textbf{p}+\textbf{q}}^{+}\hat{a}_{\textbf{k}-\textbf{q}}^{+}\hat{a}_{\textbf{k}}\hat{a}_{\textbf{p}}+\]\vspace*{-5mm}
\[
+V_{q}^{\nu_{h}\nu_{h}\nu_{h}\nu_{h}}\hat{b}_{\textbf{p}+\textbf{q}}^{+}\hat{b}_{\textbf{k}-\textbf{q}}^{+}\hat{b}_{\textbf{k}}\hat{b}_{\textbf{p}}-\]\vspace*{-5mm}
\begin{equation}-2\,V_{q}^{\nu_{e}\nu_{h}\nu_{h}\nu_{e}}\hat{a}_{\textbf{p}+\textbf{q}}^{+}\hat{b}_{\textbf{k}-\textbf{q}}^{+}\hat{b}_{\textbf{k}}\hat{a}_{\textbf{p}},
\end{equation}
where

\[
V_{q}^{\nu_{\alpha}\nu_{\beta}\nu_{\beta}\nu_{\alpha}}\!=\!\frac{e^{2}}{\kappa}\frac{1}{A}\int\limits_{-w/2}^{+w/2}dz\int\limits_{-w/2}^{+w/2}dz'\chi_{\nu_{\alpha}}(z)\chi_{\nu_{\beta}}(z')\frac{2\pi}{q}\times\]\vspace*{-5mm}
\begin{equation}
\times\,e^{-q|z-z'|}\chi_{\nu_{\beta}}(z')\chi_{\nu_{\alpha}}(z),
\end{equation}
is the Coulomb potential of the quantum well, $\kappa$ is the dielectric permittivity of a host material of the quantum well, and
$A$ is the area of the quantum well in the $xy$ plane.

The Hamiltonian of the interaction of a dipole with an electromagnetic field is described as follows:

\[
\hat{H}_{\rm
int}=-\frac{1}{A}\sum_{\nu_{e},\nu_{h},\textbf{p}}((\mu_{\textbf{p}}^{\nu_{e}\nu_{h}})^{\star}\hat{p}_{\textbf{p}}^{\nu_{e}\nu_{h}}E^{\star}e^{i\omega\,t}+\]\vspace*{-5mm}
\begin{equation}
+(\mu_{\textbf{p}}^{\nu_{e}\nu_{h}})(\hat{p}_{\textbf{p}}^{\nu_{e}\nu_{h}})^{+}Ee^{-i\omega\,t}),
\end{equation}
where
$\hat{p}_{\textbf{p}}^{\nu_{e}\nu_{h}}=\langle\,\hat{b}_{-\textbf{p}}\hat{a}_{\textbf{p}}\rangle$
is a microscopic dipole due to an electron-hole pair with the electron (hole) momentum \textbf{p} (--\textbf{p}) and the subband number
$\nu_{e}$ ($\nu_{h}$),
$\mu_{\textbf{k}}^{\nu_{e}\nu_{h}}=\int{d^{3}rU_{j'\sigma'\,\textbf{k}}\textbf{e}\hat{\mathbf{p}}U_{j\sigma\,\textbf{k}}}$,
is the matrix element of the electric dipole moment, which depends on the wave vector \textbf{k} and the numbers of subbands, between which
the direct interband transitions occur, $\textbf{e}$ is a unit vector of the vector potential of an electromagnetic wave,
$\hat{\mathbf{p}}$ is the momentum operator. Subbands are described by the wave functions $U_{j'\sigma'\,\textbf{k}}$,
$U_{j\sigma\,\textbf{k}}$, where $j'$  is the number of a subband from the conduction band, $\sigma'$ is the electron spin, $j$ is the number
of a subband from the valence band, and $\sigma$ is the hole spin. We consider one lowest conduction subband $j'=1$ and one
highest valence subband $j=1$. $E$ and $\omega$  are the electric field amplitude and frequency of an optical wave.

The polarization equation for the wurtzite quantum well in the Hartree--Fock approximation with regard for the wave functions for an
electron and a hole written in the form \cite{{Lokot},{Lokot1}}, where the coefficients of the expansion of the wave function of a hole
in the basis of wave functions (known as spherical functions) with the orbital angular momentum $l=1$ and the eigenvalue $m_{l}$ of its $z$
component, depend on the wave vector can look for as
follows:

\begin{equation}
\begin{array}{c}
\frac{d\hat{p}_{\textbf{p}}^{\nu_{e}\nu_{h}}}{dt}=-i\omega_{\textbf{p}}^{\nu_{e}\nu_{h}}\hat{p}_{\textbf{p}}^{\nu_{e}\nu_{h}}-i\Omega_{\textbf{p}}^{\nu_{e}\nu_{h}}(-1+\hat{n}_{\textbf{p}}^{\nu_{e}}+\hat{n}_{\textbf{p}}^{\nu_{h}}).\\
\end{array}
\end{equation}
The transition frequency $\omega_{\textbf{p}}^{\nu_{e}\nu_{h}}$ and
the Rabi frequency with regard for the wave function
\cite{{Lokot},{Lokot1}} are described as
\begin{equation}
\omega_{\textbf{p}}^{\nu_{e}\nu_{h}}=\frac{1}{\hbar}(\epsilon_{g0}+\epsilon_{e,\textbf{p}}^{\nu_{e}}+\epsilon_{h,\textbf{p}}^{\nu_{h}}),\end{equation}\vspace*{-5mm}

\begin{equation}
\Omega_{\textbf{p}}^{\nu_{e}\nu_{h}}=\frac{1}{\hbar}(\mu_{\textbf{p}}^{\nu_{e}\nu_{h}}Ee^{-i\omega\,t}+\sum_{\textbf{q}}V_{\scriptsize\left\{\begin{array}{c}
|-\textbf{p}|\\
|-\textbf{p}-\textbf{q}|\\
\end{array}\right\}}^{\nu_{e}\nu_{h}\nu_{h}\nu_{e}})\hat{p}_{\textbf{p}+\textbf{q}}^{\nu_{e}\nu_{h}},\\
\end{equation}
where $\epsilon_{e,\textbf{p}}^{\nu_{e}}\,,\epsilon_{h,\textbf{p}}^{\nu_{h}}$ - Hartree-Fock energies for electron and holes,

\[
V_{\scriptsize\left\{\begin{array}{c}
|-\textbf{p}|\\
|-\textbf{p}-\textbf{q}|\\
\end{array}\right\}}^{\nu_{e}\nu_{h}\nu_{h}\nu_{e}}=\frac{1}{2}\frac{e^{2}}{\kappa}\frac{1}{2\pi}\int\limits_{0}^{2\pi}d\varphi\sum_{\alpha}g_{\alpha}\int dq\times\]\vspace*{-5mm}
\[\times\int\,dz_{\xi}\int\,dz_{\xi'}\chi_{n_{1}}(z_{\xi})\chi_{m_{1}}(z_{\xi'})\chi_{m_{2}}(z_{\xi'})\chi_{n_{2}}(z_{\xi})\times\]\vspace*{-5mm}
\[
\times\,e^{-q|z_{\xi}-z_{\xi'}|}C_{p}^{j}[n_{1},1]V_{p}^{j}[m_{1},1]C_{Q_{1}}^{i}[n_{2},1]V_{Q_{1}}^{i}[m_{2},1],\]\vspace*{-7mm}
\[n_{1}=m_{1}=n_{2}=m_{2}=1,\]\vspace*{-7mm}
\begin{equation}
\textbf{Q}_{1}=\textbf{q}+\textbf{p},
\end{equation}
where $\chi_{n_{1}}(z_{\xi})$ is the envelope of the wave functions of the quantum well, $V_{p}^{i}[m_{1},1]$ and
$C_{p}^{j}[n_{1},1]$  are coefficients of the expansion of the wave functions of a hole and electron at the envelope part, $\varphi$ is
the angle between the vectors $\textbf{p}$ and $\textbf{q}$, and $g_{\alpha}$ is a degeneracy order of a level.

Numerically solving this integro-differential equation, we can obtain the absorption coefficient of a plane wave in the medium from
the Maxwell equations:

\begin{equation}
\alpha(\omega)=\frac{\omega}{\kappa\,ncE}{\rm Im}\,P,
\end{equation}
where $c$ the velocity of light in vacuum, $n$  is a background refractive index of the quantum well material,

\begin{equation}
P=\frac{2}{A}\sum_{\nu_{e},\nu_{h},\textbf{p}}(\mu_{\textbf{p}}^{\nu_{e}\nu_{h}})^{\star}p_{\textbf{p}}^{\nu_{e}\nu_{h}}e^{i\omega\,t}.
\end{equation}

The light absorption spectrum presented in the paper in Fig. 2, reflects only the strict TE (\textit{x} or \textit{y}) light polarization.

From Uncertainty Heisenberg principle:
\begin{equation}
\Delta\,x\Delta\,p\geq\,\frac{\hbar}{2},
\end{equation}

can be found the localization range particle-hole pair $\Delta\,x\geq\,\frac{\hbar}{4\,m\,c}$.

Table 1. The localization range particle-hole pair $\Delta\,x$ in cm, exciton binding energy Ry in meV, carriers concentration $n=p$ in cm$^{-2}$, Bohr radius $a_{B}$ in cm.

\begin{tabular}{ccccc} \hline\hline
\multicolumn{1}{c}{$\Delta\,x$} &
\multicolumn{1}{c}{Ry} &
\multicolumn{1}{c}{n=p} &
\multicolumn{1}{c}{$a_{B}$} & \\
\hline
$9.95*10^{-10}$ & 2.16 & $4*10^{12}$ & $4.26*10^{-6}$  \\ \hline\hline
\end{tabular}

Hence the exciton Bohr radius is grater than the localization range particle-hole pair, and the excitons may be spontaneously created.

\subsection{Results and discussions}

We consider QCSE in strained w\"{u}rtzite $\textrm{ZnO}/\textrm{Mg}_{0.27}\textrm{Zn}_{0.73}\textrm{O}$ quantum well with width 6 nm, in which the barrier height is a constant value for electrons and is equal to $U_{0}=536.22$ meV. The theoretical analysis of piezoelectric effects and exchange-correlation effects is based on the self-consistent solution of the Schr\"{o}dinger equations for electrons and holes in quantum well of width $w$ with including Stark effect and the Poisson equations. The one-dimensional Poisson equation contains the Hartree potential which includes the one-dimensional charge density for electrons and holes along the polarization field distribution. The three-dimensional Poisson equation contains besides the one-dimensional charge density for electrons and holes along the polarization field distribution the exchange-correlation potential which is built on convolutions of a plane-wave part of wave functions in addition.

We have calculated carriers population of the lowest conduction band and the both heavy hole and light hole valence band. Solving (13) for holes in the infinitely deep quantum well and finding the minimum of functional (17) for electrons in a quantum well with barriers of finite height, we can find the energy and wave functions of electrons and holes with respect to Hartree potential and exchange-correlational potential in a piezoelectric field at a carriers concentration $n=p=4*10^{12}$ $\textrm{cm}^{-2}$. The screening field is determined by iterating Eqs. (13), (17), (22) until the solution of energy spectrum is converged.

The Hartree-Fock dispersions of the valence bands and the conduction band are presented in Fig. 1. The light absorption spectrum presented in the paper in Fig. 2.

It is found that the localization range particle-hole pair $\Delta\,x\geq\,\frac{\hbar}{4\,m\,c}\sim\,9.95*10^{-10}$ cm. Exciton binding energy is equal Ry=2.16 meV at carriers concentration $n=p=4*10^{12}$ cm$^{-2}$. Bohr radius is equal $a_{B}=4.26*10^{-6}$ cm.

If the exciton Bohr is grater than the localization range particle-hole pair, the excitons may be spontaneously created.

We consider the pairing between oppositely charged particles with complex dispersion. The Coulomb interaction leads to the electron-hole bound states scrutiny study of which acquire significant attention in the explanations of high-temperature superconductivity. If the exciton Bohr radius is grater than the localization range particle-hole pair, the excitons may be spontaneously created.

It is found that $E_{(Hartree-Fock\,band\,gap)}-E_{(1s\,exciton)}=0.2$ meV. If the electron and hole are separated, their energy is higher on 0.2 meV than if they are paired. Hence it can be energetically favorable for them to be paired.

If the Hartree-Fock band gap energy is greater than the exciton energy in ZnO/(Zn,Mg)O quantum wells then excitons may be spontaneously created. It is known in narrow-gap semiconductor or semimetal then at sufficiently low temperature the insulator ground state is instable with respect to the exciton formation ~\cite{{Stroucken},{Jerome}}, leading to a spontaneously creating of excitons. In a system undergo a phase transition into a exciton insulator phase similarly to Bardeen-Cooper-Schrieffer (BCS) superconductor.

An exciton insulator states with a gap 3.4 eV are predicted. The particle-hole pairing leads to the Cooper instability.

\section{Elliott formula for particle-hole pair of Dirac cone.}

The graphene ~\cite{{Novoselov1},{Novoselov2},{Vasko}} presents a new state of matter of layered materials. The energy bands for graphite was found using "tight-binding" approximation by P.R. Wallace ~\cite{{Wallace}}. In the low-energy limit the single-particle spectrum is Dirac cone similarly to the light cone in relativistic physics, where the light velocity is substituted by the Fermi velocity $v_{F}$ and describes by the massless Dirac equation.

In the paper we present a theoretical investigation of excitonic states as well as their wave functions in graphene. An integral form of the two-dimensional Schr\"{o}dinger equation of Kepler problem in momentum space is solved exactly by projection the two-dimensional space of momentum on the three-dimensional sphere in the paper ~\cite{{Parfitt}}.

The integral Schr\"{o}dinger equation was analytically solved by the projection the three-dimensional momentum space onto the surface of a four-dimensional unit sphere by Fock in 1935 ~\cite{{Fock}}.

We consider the pairing between oppositely charged particles with complex dispersion. The Coulomb interaction leads to the electron-hole bound states scrutiny study of which acquire significant attention in the explanations of superconductivity.

If the exciton binding energy is greater than the flat band gap in narrow-gap semiconductor or semimetal then at sufficiently low temperature the insulator ground state is instable with respect to the exciton formation ~\cite{{Stroucken},{Jerome}}. And excitons may be spontaneously created. In a system undergo a phase transition into a exciton insulator phase similarly to Bardeen-Cooper-Schrieffer (BCS) superconductor. In a single-layer graphene (SLG) the electron-hole pairing leads to the exciton insulator states ~\cite{{Lokot}}.

In the paper an integral two-dimensional Schr\"{o}dinger equation of the electron-hole pairing for particles with complex dispersion is analytically solved. A complex dispersions lead to fundamental difference in exciton insulator states and their wave functions.

We analytically solve an integral two-dimensional Schr\"{o}dinger equation of the electron-hole pairing for particles with electron-hole symmetry of reflection.

For graphene in vacuum the effective fine structure parameter $\alpha_{G}=\frac{e^{2}}{v_{F}\hbar\kappa\sqrt{\pi}}=1.23$. For graphene in substrate $\alpha_{G}=0.77$, when the permittivity of graphene in substrate is estimated to be $\kappa=1.6$ ~\cite{{Alicea}}. It means the prominent Coulomb effects ~\cite{{Sharapov}}.

\begin{figure}
\includegraphics*[bb=5 10 1000 600,width=5in]{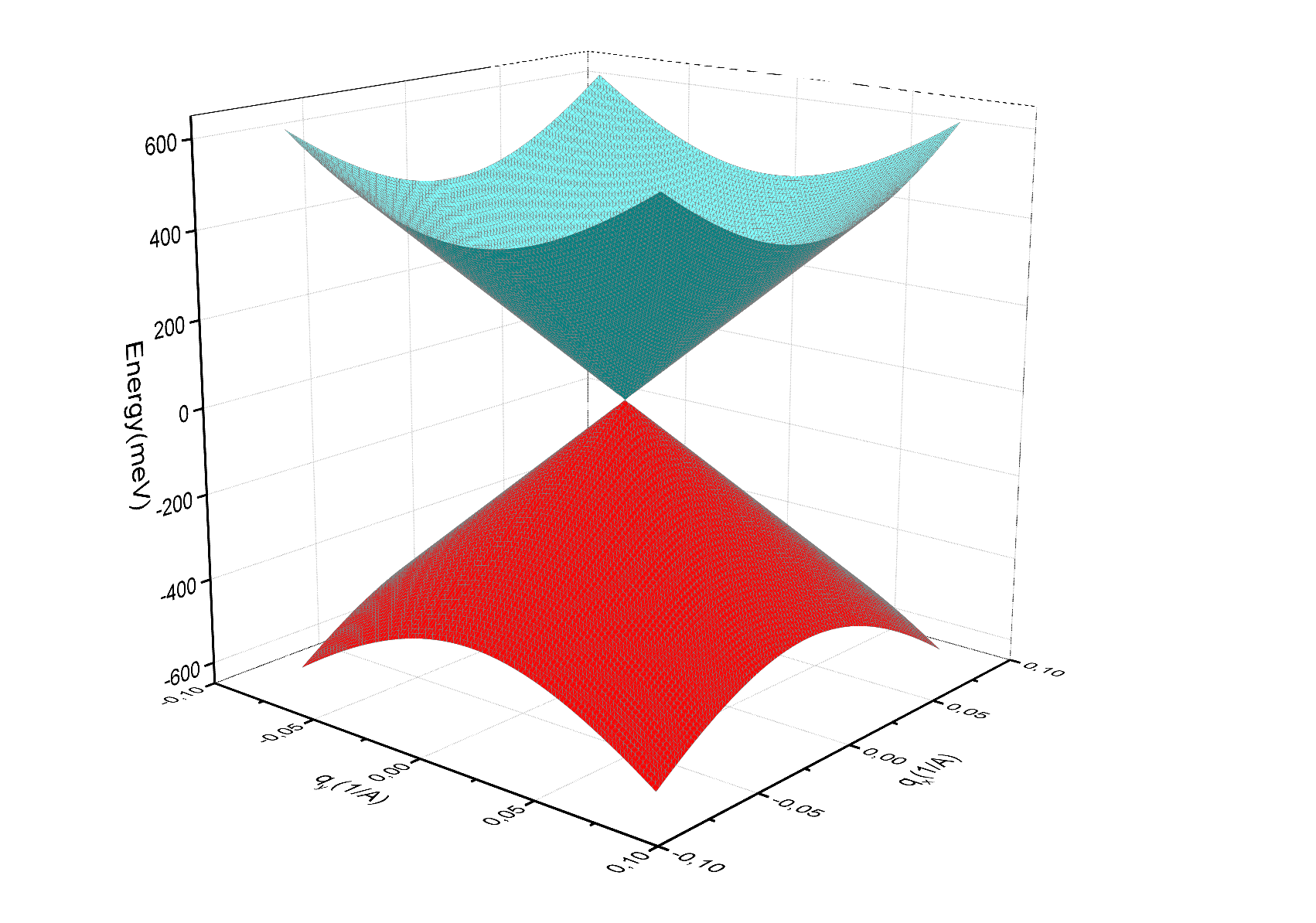}
\caption{(Color online) Single-particle spectrum of graphene for massless Dirac fermions (Majorana fermions).}
\end{figure}

It is known that the Coulomb interaction leads to the semimetal-exciton insulator transition, where  gap is opened by electron-electron exchange interaction ~\cite{{Jerome},{Stroucken1},{Kadi},{Malic}}. The perfect host combines a small gap and a large exciton binding energy ~\cite{{Jerome},{Stroucken}}.

In graphene the existing of bound pair states are still subject matter of researches ~\cite{{Gamayun},{Gamayun1},{Berman},{Berman1},{Hartmann}}.

It is known ~\cite{{Min}} in the weak-coupling limit ~\cite{{Sak}}, exciton condensation is a consequence of the Cooper instability of materials with electron-hole symmetry of reflection inside identical Fermi surface. The identical Fermi surfaces is a consequence of the particle-hole symmetry of massless Dirac equation for Majorana fermions.

\begin{widetext}

\subsection{Quantized spectral series of the excitonic states of valence Dirac cone.}

In the honeycomb lattice of graphene with two carbon atoms per unit cell the space group is $D_{3h}^{1}$ ~\cite{{Malard}}:

\begin{tabular}{cccccccc} \hline\hline
\multicolumn{1}{c}{$D_{3h}^{1}$} &
\multicolumn{1}{c}{${\{E|0\}}$} &
\multicolumn{1}{c}{${\{C^{(+,-)}_{3}|0\}}$} &
\multicolumn{1}{c}{${\{C_{2}'^{(A,B,C)}|0\}}$} &
\multicolumn{1}{c}{${\{\sigma_{h}|\tau\}}$} &
\multicolumn{1}{c}{${\{S^{(-,+)}_{3}|\tau\}}$} &
\multicolumn{1}{c}{${\{\sigma^{(A,B,C)}_{v}|\tau\}}$} &
\multicolumn{1}{c}{$ $} \\
\hline
$K_{3}^{+}$ & 2 & -1 & 0 & 2 & -1 & 0 & $ $ \\
$g^{2}$ & ${\{E|0\}}$ & ${\{C^{(+,-)}_{3}|0\}}$ & ${\{E|0\}}$ & ${\{E|0\}}$ & ${\{S^{(-,+)}_{3}|\tau\}}$ & ${\{E|0\}}$ & $ $ \\
$\chi^{2}(g)$ & 4 & 1 & 0 & 4 & 1 & 0 & $K^{+}_{1}+K^{+}_{2}+K^{+}_{3}$\\
$\chi(g^{2})$ & 2 & -1 & 2 & 2 & -1 & 2 & $ $\\
$\frac{1}{2}[\chi^{2}(g)+\chi(g^{2})]$ & 3 & 0 & 1 & 3 & 0 & 1 & $K^{+}_{1}+K^{+}_{3}$\\
$\frac{1}{2}\{\chi^{2}(g)-\chi(g^{2})\}$ & 1 & 1 & -1 & 1 & 1 & -1 & $K^{+}_{2}$\\ \hline\hline
\end{tabular}
\end{widetext}

The direct production of two irreducible presentations of wave function and wave vector of difference $\kappa-K$ or $\kappa-K'$ expansion is $K^{+}_{3}\times\,K_{3}^{+\star}$ and can be expanded on

\begin{equation}
p^{\alpha}: \tau_{\psi}\times\,\tau_{k}=(K^{+}_{1}+K^{+}_{2}+K^{+}_{3})\times\,K^{+}_{3}=K^{+}_{3}\times\,K^{+}_{3}.
\end{equation}

In the low-energy limit the single-particle spectrum is Dirac cone describes of the massless Dirac equation for a massless Dirac fermions (Majorana fermions). The Hamiltonian of graphene for a massless Dirac fermions ~\cite{{Wallace}}

\begin{equation}
\hat{H}=v_{F}(\tau\,q_{x}\hat{\sigma_{x}}+q_{y}\hat{\sigma_{y}}),
\end{equation}
where
$q_{x}$, $q_{y}$ are Cartesian components of a wave vector, $\tau=\pm\,1$ is the valley index, $v_{F}=10^{6}$ m/s is the graphene Fermi velocity, $\hat{\sigma_{x}}$, $\hat{\sigma_{y}}$ are Pauli matrices (here we assume that $\hbar=1$).

The dispersion of energy bands may be found in the form ~\cite{{Wallace}}
\begin{equation}
\begin{array}{c}
\epsilon_{\pm}=\pm\,v_{F}q,\\
\end{array}
\end{equation}
where
$q=\sqrt{q_{x}^{2}+q_{y}^{2}}$.

The Schr\"{o}dinger equation for the calculating of exciton states can be written in the general form
\begin{equation}
(\epsilon(q)+q_{0}^{2})\Phi(\textbf{q})=\frac{1}{\pi}\int{\frac{\Phi(\textbf{q}')}{|\textbf{q}-\textbf{q}'|}}d\textbf{q}',
\end{equation}
where $q_{0}^{2}=-\epsilon$, $\epsilon$ is a quantized energy. We look for the bound states and hence the energy will be negative.

For the single layer graphene
\begin{equation}
\begin{array}{c}
\frac{\epsilon(q)+q_{0}^{2}}{q^{2}+q_{0}^{2}}=\pm\,\frac{v_{F}}{2q_{0}}\sin{(\theta)}+\frac{1-\cos{\theta}}{2}.
\end{array}
\end{equation}

An integral form of the two-dimensional Schr\"{o}dinger equation in momentum space for the graphene is solved exactly by projection the two-dimensional space of momentum on the three-dimensional sphere.

When an each point on sphere is defined of two spherical angles $\theta$, $\phi$, which are knitted with a momentum $\textbf{q}$ ~\cite{{Fock},{Parfitt}}. A space angle $\Omega$ may be found as surface element on sphere $d\Omega=\sin({\theta})d\theta\,d\phi=(\frac{2q_{0}}{q^{2}+q_{0}^{2}})^{2}d\textbf{q}$ ~\cite{{Fock},{Parfitt}}.
A spherical angle $\theta$ and a momentum $\textbf{q}$ are shown ~\cite{{Fock},{Parfitt}} to be knitted as
\begin{equation}
\cos{\theta}=\frac{q^{2}-q_{0}^{2}}{q^{2}+q_{0}^{2}},\,\sin{\theta}=\frac{2qq_{0}}{q^{2}+q_{0}^{2}},\,q^{2}=q_{0}^{2}(\frac{1+\cos{\theta}}{1-\cos{\theta}}).
\end{equation}

Using spherical symmetry the solution of integral Schr\"{o}dinger equation can look for in the form
\begin{equation}
\Phi(\textbf{q})=\sqrt{q_{0}}(\frac{2q_{0}}{q^{2}+q_{0}^{2}})^{3/2}\sum_{l=0}^{\infty}A_{l}Y_{l}^{0}(\theta,\phi),
\end{equation}
where
\begin{equation}
\begin{array}{c}
Y_{l}^{0}(\theta,\phi)=\sqrt{\frac{2l+1}{4\pi}}P_{l}^{0}(\cos{\theta}).
\end{array}
\end{equation}

Since ~\cite{{Parfitt}}
\begin{widetext}
\begin{equation}
\begin{array}{c}
\frac{(q^{2}+q_{0}^{2})^{1/2}(q'^{2}+q_{0}^{2})^{1/2}}{2q_{0}}\frac{1}{|\textbf{q}-\textbf{q}'|}=\sum_{\lambda=0}^{\infty}\sum_{\mu=-\lambda}^{\lambda}\frac{4\pi}{2\lambda+1}Y_{\lambda}^{\mu}(\theta,\phi)Y_{\lambda}^{\mu,\ast}(\theta',\phi'),
\end{array}
\end{equation}
then substituting (65), (67) in (62), can find equation
\begin{equation}
\begin{array}{c}
\frac{\epsilon(q)+q_{0}^{2}}{q^{2}+q_{0}^{2}}\sum_{l=0}^{\infty}A_{l}Y_{l}^{0}(\theta,\phi)=\frac{2}{q_{0}}\sum_{l=0}^{\infty}\sum_{\lambda=0}^{\infty}\sum_{\mu=-\lambda}^{\lambda}\int\frac{1}{2\lambda+1}Y_{\lambda}^{\mu}(\theta,\phi)Y_{\lambda}^{\mu,\ast}(\theta',\phi')Y_{l}^{0}(\theta',\phi')A_{l}(\frac{2q_{0}}{q'^{2}+q_{0}^{2}})^{2}d\textbf{q}'.
\end{array}
\end{equation}

The integral equations for SLG based on Eq. (63) may be found in the form
\begin{equation}
\begin{array}{c}
\int(\pm\,\frac{v_{F}}{2q_{0}}\sin{(\theta)}+\frac{1-\cos{\theta}}{2})\sum_{l=0}^{\infty}A_{l}Y_{l}^{0}(\theta,\phi)Y_{k}^{n,\ast}(\theta,\phi)d\Omega=\\
=\frac{2}{q_{0}}\int\sum_{\lambda=0}^{\infty}\sum_{\mu=-\lambda}^{\lambda}\sum_{l'=0}^{\infty}\frac{1}{2\lambda+1}Y_{\lambda}^{\mu}(\theta,\phi)Y_{\lambda}^{\mu,\ast}(\theta',\phi')Y_{l'}^{0}(\theta',\phi')Y_{k}^{n,\ast}(\theta,\phi)d\Omega\,d\Omega'A_{l'}.
\end{array}
\end{equation}

Since ~\cite{{Fock1}}
\begin{equation}
\begin{array}{c}
\cos{\theta}P_{l}^{m}(\cos{\theta})=\frac{\sqrt{l^{2}-m^{2}}}{\sqrt{4l^{2}-1}}P_{l-1}^{m}(\cos{\theta})+\frac{\sqrt{(l+1)^{2}-m^{2}}}{\sqrt{4(l+1)^{2}-1}}P_{l+1}^{m}(\cos{\theta}),
\end{array}
\end{equation}
\begin{equation}
\begin{array}{c}
\sin{\theta}P_{l}^{m}(\cos{\theta})=\frac{\sqrt{(l-m)(l-m-1)}}{\sqrt{4l^{2}-1}}P_{l-1}^{m+1}(\cos{\theta})+\frac{\sqrt{(l+m+1)(l+m+2)}}{\sqrt{4(l+1)^{2}}-1}P_{l+1}^{m+1}(\cos{\theta}),
\end{array}
\end{equation}
\end{widetext}
then solutions of the integral equation (68) for the energies and wave functions correspondingly can be found analytically with taken into account the normalization condition $(\frac{1}{2\pi})^2\int{\frac{q^{2}+q_{0}^{2}}{2q_{0}^{2}}|\Phi(\textbf{q})|^{2}d\textbf{q}}=1$.

From equation (69) one can obtain the eigenvalue and eigenfunction problem
one can find recurrence relation
\begin{equation}
\frac{1}{2}(l+\frac{1}{2})A_{l}+\frac{1}{q_{0}}A_{l}+\frac{1}{2}A_{l-1}(l+\frac{1}{2})a_{l}+\frac{1}{2}A_{l+1}(l+\frac{1}{2})b_{l}=0.
\end{equation}

The solutions of the quantized series in excitonic Rydbergs where Ry=$87.37$ meV, and wave functions of the integral equation (69) one can find in the form
\begin{equation}
\epsilon_{0}=-\frac{1}{(\frac{1}{4}+\frac{1}{2}(1+\frac{1}{2})a_{1})^{2}},
\end{equation}
\begin{equation}
\epsilon_{1}=-\frac{1}{(\frac{1}{2}(1+\frac{1}{2})+\frac{1}{4}b_{0}+\frac{1}{2}(2+\frac{1}{2})a_{2})^{2}},
\end{equation}
\begin{equation}
\epsilon_{2}=-\frac{1}{(\frac{1}{2}(2+\frac{1}{2})+\frac{1}{2}(1+\frac{1}{2})b_{1}+\frac{1}{2}(3+\frac{1}{2})a_{3})^{2}},
\end{equation}
\begin{equation}
\epsilon_{3}=-\frac{1}{(\frac{1}{2}(3+\frac{1}{2})+\frac{1}{2}(2+\frac{1}{2})b_{2}+\frac{1}{2}(4+\frac{1}{2})a_{4})^{2}},
\end{equation}
\begin{equation}
\begin{array}{c}
\Phi_{l}(\cos{\theta})=\sqrt{\frac{2\pi}{(q_{0l})^3}}\sum_{n=0}^{\infty}(1-\cos{\theta})^{3/2}P_{n}^{0}(\cos{\theta}),
\end{array}
\end{equation}
where $q_{0l}^{2}=-\epsilon_{l}$, $l=0,1,2,3,4,....$,
\begin{equation}
a_{l}=\frac{1}{2\pi}\sqrt{\frac{2(l-1)+1}{2}}\sqrt{\frac{2}{2l+1}}\frac{l}{\sqrt{4l^{2}+1}},
\end{equation}
\begin{equation}
b_{l}=\frac{1}{4\pi}\sqrt{2(l+1)+1}\sqrt{2l+1}\frac{l+1}{\sqrt{4(l+1)^{2}-1}}.
\end{equation}

Table 2. Quantized spectral series of the excitonic states which distribute in valence cone $\epsilon_{n}$, $n=0,1,2,3,...$ in meV, exciton Rydberg Ry in meV.

\begin{tabular}{cccccccc} \hline\hline
\multicolumn{1}{c}{$\epsilon_{0}$} &
\multicolumn{1}{c}{$\epsilon_{1}$} &
\multicolumn{1}{c}{$\epsilon_{2}$} &
\multicolumn{1}{c}{$\epsilon_{3}$} &
\multicolumn{1}{c}{Ry} \\
\hline
1107.94& 122.47 & 39.59 & 17.97 & 87.37 \\ \hline\hline
\end{tabular}

Quantized spectral series of the excitonic states distribute in valence Dirac cone. The energies of bound states are shown to be found as negative, i. e. below of Fermi level.  Thus if the electron and hole are separated, their energy is higher than if they are paired.

\subsection{Elliott formula and light absorption rates of the excitonic states of valence Dirac cone.}

The intervalley transitions probability caused intervalley photoexcitations taken into account Coulomb interaction of electron-hole pair one can obtain from Fermi golden rule in the form

\begin{equation}
\begin{array}{c}
P=\frac{2\pi}{\hbar}(\frac{ev_{F}E_{\omega}}{\hbar\omega})^{2}\sum_{n}(\sum_{\textbf{q}}|\langle\,\pm\,1,\textbf{q}|\hat{\sigma}_{x,y}|\mp\,1,\textbf{q}\rangle|\times\\
\times\Phi_{n}(\frac{q^{2}-q_{0}^{2}}{q^{2}+q_{0}^{2}}))^{2}[\delta_{\gamma}(\epsilon_{n}-\hbar\,\omega)+\delta_{\gamma}(\epsilon_{n}+\hbar\,\omega)].\\
\end{array}
\end{equation}

Considering the case of relatively weak excitation the total rate of increase of the number of photons in the fixed mode one can obtain in the form

\begin{equation}
\begin{array}{c}
R=\frac{2\pi}{\hbar}(\frac{ev_{F}E_{\omega}}{\hbar\omega})^{2}\sum_{n}(\sum_{\textbf{q}}|\langle\,\pm\,1,\textbf{q}|\hat{\sigma}_{x,y}|\mp\,1,\textbf{q}\rangle|\times\\
\times\Phi_{n}(\frac{q^{2}-q_{0}^{2}}{q^{2}+q_{0}^{2}}))^{2}[\delta_{\gamma}(\epsilon_{n}-\hbar\,\omega)+\delta_{\gamma}(\epsilon_{n}+\hbar\,\omega)].\\
\end{array}
\end{equation}

The change in the energy density of electromagnetic waves can be presented in the form

\begin{equation}
\frac{dW}{dt}=\frac{\hbar\,\omega}{S}R.
\end{equation}

Under ac electric field $E_{\omega}\textbf{e}\cos{(qz-\omega\,t)}$ the energy density of electromagnetic waves one can obtain in the form $W=\frac{1}{8\pi}\kappa\,E_{\omega}^{2}$. Light absorption rate one can obtain in the form $\alpha(\omega)=\frac{1}{W}\frac{dW}{dz}$. Since $\frac{dW}{dz}=\frac{dW}{dt}\frac{\sqrt{\kappa}}{c}$ then light absorption rate with taken into account $|\langle\,\pm\,1,\textbf{q}|\hat{\sigma}_{x,y}|\mp\,1,\textbf{q}\rangle|^{2}=1/2$ can be rewritten in the form

\begin{equation}
\begin{array}{c}
\Im(\alpha(\omega))=\frac{16}{\sqrt{\kappa}c\hbar^{2}\omega}(ev_{F})^{2}\sum_{n}(\int\,d\textbf{q}\times\\
\times\Phi_{n}(\frac{q^{2}-q_{0}^{2}}{q^{2}+q_{0}^{2}}))^{2}[\delta_{\gamma}(\epsilon_{n}-\hbar\,\omega)+\delta_{\gamma}(\epsilon_{n}+\hbar\,\omega)],\\
\end{array}
\end{equation}

where $\sum_{\textbf{q}}\rightarrow\,\frac{S}{(2\pi)^{2}}\int\,d\textbf{q}$ in a formula (81).

Table 3. Light absorption rate of quantized spectral series of the excitonic states which distribute in valence cone $\alpha_{n}$, $n=0,1,...$ in cm$^{-1}$.

\begin{tabular}{ccc} \hline\hline
\multicolumn{1}{c}{$\alpha_{0}$} &
\multicolumn{1}{c}{$\alpha_{1}$} & \\
\hline
7.67*10$^{22}$ & 1.14*10$^{25}$ \\ \hline\hline
\end{tabular}

\subsection{Results and discussions}

The integral Schr\"{o}dinger equation for a parabolic bands was analytically solved by the projection the three-dimensional momentum space onto the surface of a four-dimensional unit sphere by Fock in 1935 ~\cite{{Fock}}.

In the paper an integral two-dimensional Schr\"{o}dinger equation of the electron-hole pairing for particles with complex dispersion is analytically solved. A complex dispersion leads to fundamental difference in the energy of exciton insulator states and their wave functions.

We analytically solve an integral two-dimensional Schr\"{o}dinger equation of the electron-hole pairing for particles with electron-hole symmetry of reflection.

It is known that the Coulomb interaction leads to the semimetal-exciton insulator transition, where  gap is opened by electron-electron exchange interaction ~\cite{{Jerome},{Stroucken1},{Kadi},{Malic}}. The perfect host combines a small gap and a large exciton binding energy ~\cite{{Jerome},{Stroucken}}.

We consider the pairing between oppositely charged particles in graphene. The Coulomb interaction leads to the electron-hole bound states scrutiny study of which acquire significant attention in the explanations of superconductivity.

It is known ~\cite{{Stroucken},{Jerome}} if the exciton binding energy is greater than the flat band gap in narrow-gap semiconductor or semimetal then at sufficiently low temperature the insulator ground state is instable concerning to the exciton formation with follow up spontaneous production of excitons. In a system undergo a phase transition into a exciton insulator phase similarly to BCS superconductor. In a SLG the electron-hole pairing leads to the exciton insulator states.

The particle-hole symmetry of Dirac equation of layered materials allows perfect pairing between electron Fermi sphere and hole Fermi sphere in the valence band and conduction band and hence driving the Cooper instability. In the weak-coupling limit in graphene with the occupied conduction-band states and empty valence-band states inside identical Fermi surfaces in band structure, the exciton condensation is a consequence of the Cooper instability.

\section{Conclusions}

In this paper a theoretical studies of the space separation of electron and hole wave functions in the quantum well $\textrm{ZnO}/\textrm{Mg}_{0.27}\textrm{Zn}_{0.73}\textrm{O}$ by the self-consistent solution of the Schr\"{o}dinger equations for electrons and holes and the Poisson equations at the presence of spatially varying quantum well potential due to the piezoelectric effect and local exchange-correlation potential are presented. The exchange-correlation potential energy is found from the solution of three-dimensional Poisson's equation, using both an expression by Gunnarsson and Lundquist ~\cite{{Gunnarsson}}, and following criterions. The criterion $k_{F}>\sqrt{n}/4$ at carrier densities $4*10^{12}$ $\textrm{cm}^{-2}$, at a temperature T=0 K is carried as $1>0.1$. The criterion does not depend from a width of well. The solution of equations system (13), (17), (23) as well as (13), (17), (22) does not depend from a temperature. The ratio of Coulomb potential energy to the Fermi energy is $r_{s}=E_{C}/E_{F}=0.63<1$. The one-dimensional Poisson equation contains the Hartree potential which includes the one-dimensional charge density for electrons and holes along the polarization field distribution. The three-dimensional Poisson equation contains besides the one-dimensional charge density for electrons and holes along the polarization field distribution the exchange-correlation potential which is built on convolutions of a plane-wave part of wave functions in addition. The problem consists of the one-dimensional Poisson's equation solving of which may be found Hartree potential energy and three-dimensional Poisson's equation which is separated on one-dimensional and two-dimensional equations by separated of variables. At the condition that the ratio of wave function localization in the longitudinal z direction on transversal in-plane wave function localization is less 1. It is found that the localization range particle-hole pair $\Delta\,x\geq\,\frac{\hbar}{4\,m\,c}\sim\,9.95*10^{-10}$ cm. Exciton binding energy is equal Ry=2.16 meV at carriers concentration $n=p=4*10^{12}$ cm$^{-2}$. Bohr radius is equal $a_{B}=4.26*10^{-6}$ cm. It is found that the exciton binding energy is grater than the localization range particle-hole pair, and the excitons may be spontaneously created. If the electron and hole are separated, their energy is higher on 0.2 meV than if they are paired. Hence it can be energetically favorable for them to be paired. An exciton insulator states with a gap 3.4 eV are predicted. The particle-hole pairing leads to the Cooper instability.

In this paper we found the solution the integral Schr\"{o}dinger equation in a momentum space of two interacting via a Coulomb potential Dirac particles that form the exciton in graphene.

In low-energy limit this problem is solved analytically. We obtained the energy dispersion and wave function of the exciton in graphene. The excitons were considered as a system of two oppositely charge Dirac particles interacting via a Coulomb potential.

We solve this problem in a momentum space because on the whole the center-of-mass and the relative motion of the two Dirac particles can not be separated.

We analytically solve an integral two-dimensional Schr\"{o}dinger equation of the electron-hole pairing for particles with electron-hole symmetry of reflection. An integral form of the two-dimensional Schr\"{o}dinger equation in momentum space for graphene is solved exactly by projection the two-dimensional space of momentum on the three-dimensional sphere.

Quantized spectral series of the excitonic states distribute in valence Dirac cone. The energies of bound states are shown to be found as negative, i. e. below of Fermi level.  Thus if the electron and hole are separated, their energy is higher than if they are paired. In the SLG the electron-hole pairing leads to the exciton insulator states.

\section{Appendix A}
\subsection{Matrix elements of interband transitions}

\begin{widetext}
\begin{tabular}{cccccccc} \hline\hline
\multicolumn{1}{c}{$\Gamma_{1}\times(\Gamma_{1}+\Gamma_{5})$} &
\multicolumn{1}{c}{$E$} &
\multicolumn{1}{c}{$C_{2}$} &
\multicolumn{1}{c}{$2C_{3}$} &
\multicolumn{1}{c}{$2C_{6}$} &
\multicolumn{1}{c}{$3\sigma_{v}$} &
\multicolumn{1}{c}{$3\sigma_{v}'$} &
\multicolumn{1}{c}{$ $} \\
\hline
$\chi_{\mu}^{\psi\,\star}(g)\chi_{\nu}^{\psi}(g)$ & 3 & -1 & 0 & 2 & 1 & 1 & $\Gamma_{1}+\Gamma_{5}$ \\
$\chi_{v}$ & 3 & -1 & 0 & 2 & 1 & 1 & $\Gamma_{1}+\Gamma_{5}$ \\
\hline\hline
\end{tabular}

Matrix elements of interband transitions transforms according to representations:
\begin{equation}
\begin{array}{c}
M_{v-c}(\textbf{k}): \tau_{v}\times\,\tau_{\psi}=(\Gamma_{1}+\Gamma_{5})\times\,(\Gamma_{1}+\Gamma_{5})=\\
=\Gamma_{1}\times\Gamma_{1}+\Gamma_{5}\times\Gamma_{5}.
\end{array}
\end{equation}
\begin{tabular}{cccccccc} \hline\hline
\multicolumn{1}{c}{$C_{6v}$} &
\multicolumn{1}{c}{$E$} &
\multicolumn{1}{c}{$C_{2}$} &
\multicolumn{1}{c}{$2C_{3}$} &
\multicolumn{1}{c}{$2C_{6}$} &
\multicolumn{1}{c}{$3\sigma_{v}$} &
\multicolumn{1}{c}{$3\sigma_{v}'$} &
\multicolumn{1}{c}{$ $} \\
\hline
$\Gamma_{1}$ & 1 & 1 & 1 & 1 & 1 & 1 & $k_{z}^{2}\,,k_{t}^{2}\,,J_{z}^{2}\,,I $ \\
$\Gamma_{2}$ & 1 & 1 & 1 & -1 & -1 & -1 & $J_{z}\,,\sigma_{z} $ \\
$\Gamma_{3}$ & 1 & 1 & -1 & 1 & 1 & -1 & $ $\\
$\Gamma_{4}$ & 1 & 1 & -1 & -1 & -1 & 1 & $ $\\
$\Gamma_{5}$ & 2 & -1 & 0 & -2 & 1 & 0 & $k_{+}\,,k_{-}\,,\sigma_{+}\,,\sigma_{-}\,,J_{+}\,,J_{-}\,,[J_{+}J_{z}]\,,[J_{-}J_{z}] $\\
$\Gamma_{6}$ & 2 & -1 & 0 & 2 & -1 & 0 & $k_{+}^{2}\,,k_{-}^{2}\,,J_{+}^{2}\,,J_{-}^{2} $\\ \hline\hline
\end{tabular}
\end{widetext}
where
$k_{\pm}=k_{x}\pm\,ik_{y}$, $k_{t}^{2}=k_{x}^{2}+k_{y}^{2}$, $J_{\pm}=\frac{1}{\sqrt{2}}(J_{x}\pm\,iJ_{y})$, $2[J_{z}J_{\pm}]=J_{z}J_{\pm}+J_{\pm}J_{z}$, $\sigma_{\pm}=\frac{1}{2}(\sigma_{x}\pm\,\sigma_{y})$,

\begin{equation}
J_{+}=\left\|
\begin{array}{cccc}
0 & 1 & 0\\
0 & 0 & 1\\
0 & 0 & 0\\
\end{array}
\right\|,
\end{equation}
\begin{equation}
J_{-}=\left\|
\begin{array}{cccc}
0 & 0 & 0\\
1 & 0 & 0\\
0 & 1 & 0\\
\end{array}
\right\|,
\end{equation}
\begin{equation}
J_{z}=\left\|
\begin{array}{cccc}
1 & 0 & 0\\
0 & 0 & 0\\
0 & 0 & -1\\
\end{array}
\right\|,
\end{equation}
\begin{equation}
J_{z}^{2}=\left\|
\begin{array}{cccc}
1 & 0 & 0\\
0 & 0 & 0\\
0 & 0 & 1\\
\end{array}
\right\|,
\end{equation}
\begin{equation}
J_{+}^{2}=\left\|
\begin{array}{cccc}
0 & 0 & 1\\
0 & 0 & 0\\
0 & 0 & 0\\
\end{array}
\right\|,
\end{equation}
\begin{equation}
J_{-}^{2}=\left\|
\begin{array}{cccc}
0 & 0 & 0\\
0 & 0 & 0\\
1 & 0 & 0\\
\end{array}
\right\|,
\end{equation}
\begin{equation}
2[J_{z}J_{+}]=\left\|
\begin{array}{cccc}
0 & 1 & 0\\
0 & 0 & -1\\
0 & 0 & 0\\
\end{array}
\right\|,
\end{equation}
\begin{equation}
2[J_{z}J_{-}]=\left\|
\begin{array}{cccc}
0 & 0 & 0\\
1 & 0 & 0\\
0 & -1 & 0\\
\end{array}
\right\|,
\end{equation}
\begin{equation}
\sigma_{z}=\left\|
\begin{array}{cccc}
1 & 0\\
0 & -1\\
\end{array}
\right\|,
\end{equation}
\begin{equation}
\sigma_{+}=\left\|
\begin{array}{cccc}
0 & 1\\
0 & 0\\
\end{array}
\right\|,
\end{equation}
\begin{equation}
\sigma_{-}=\left\|
\begin{array}{cccc}
0 & 0\\
1 & 0\\
\end{array}
\right\|.
\end{equation}

\section{Appendix B}

Table 4. The irreducible representational of $D^{1}_{3h}$ ~\cite{Mildred}.
\begin{widetext}
\begin{tabular}{cccccccc} \hline\hline
\multicolumn{1}{c}{$D^{1}_{3h}$} &
\multicolumn{1}{c}{${\{E|0\}}$} &
\multicolumn{1}{c}{${\{C^{(+,-)}_{3}|0\}}$} &
\multicolumn{1}{c}{${\{C_{2}'^{(A,B,C)}|0\}}$} &
\multicolumn{1}{c}{${\{\sigma_{h}|\tau\}}$} &
\multicolumn{1}{c}{${\{S^{(-,+)}_{3}|\tau\}}$} &
\multicolumn{1}{c}{${\{\sigma^{(A,B,C)}_{v}|\tau\}}$} &
\multicolumn{1}{c}{$ $} \\
\hline
$K_{1}^{+}$ & 1 & 1 & 1 & 1 & 1 & 1 & $x^{2}+y^{2},\,z^{2}$ \\
$K_{2}^{+}$ & 1 & 1 & -1 & 1 & 1 & -1 & $J_{z}$ \\
$K_{3}^{+}$ & 2 & -1 & 0 & 2 & -1 & 0 & $(x,\,y)$\\
$K_{1}^{-}$ & 1 & 1 & 1 & -1 & -1 & -1 & $$\\
$K_{2}^{-}$ & 1 & 1 & -1 & -1 & -1 & 1 & $z$\\
$K_{3}^{-}$ & 2 & -1 & 0 & -2 & 1 & 0 & $(x^{2}-y^{2}, xy),\,(J_{x},J_{y})$\\ \hline\hline
\end{tabular}

\section{Appendix C}

From a trigonometric calculation one can find a following recurrence relations

\begin{equation}
\begin{array}{c}
\cot{\theta}P_{l}^{m+1}(\cos{\theta})=\frac{P_{l}^{m+2}(\cos{\theta})+[l(l+1)-m(m+1)]P_{l}^{m}(\cos{\theta})}{2(m+1)},
\end{array}
\end{equation}

\begin{equation}
\begin{array}{c}
\frac{1}{\sin{\theta}}P_{l-1}^{m}(\cos{\theta})=\frac{(2l+1)P_{l}^{m+1}(\cos{\theta})+(l-m)(l-m+1)(2l+1)P_{l}^{m-1}(\cos{\theta})}{((l+m)(l+m+1)-(l-m)(l-m+1))},
\end{array}
\end{equation}

\begin{equation}
\begin{array}{c}
\cot{\theta}P_{l}^{m}(\cos{2\theta})=(\frac{1}{\sin{2\theta}}+\cot{2\theta})P_{l}^{m}(\cos{2\theta}),\\
\end{array}
\end{equation}

\begin{equation}
\begin{array}{c}
(\frac{1}{2}(3+4(\cot{\theta})^{2})-\frac{1}{2}-\frac{1}{\sin{\theta}})P_{l}^{m}(\cos{2\theta})=\cot{\theta}P_{l}^{m}(\cos{2\theta}),
\end{array}
\end{equation}

where

\begin{equation}
\begin{array}{c}
P_{l}^{m}(x)=\frac{1}{2^{m}}\frac{(l+m)!}{(l-m)!m!}(1-x^{2})^{m/2}F(m-l,m+l+1,m+1,\frac{1-x}{2}),\\
\end{array}
\end{equation}

\begin{equation}
\begin{array}{c}
F(\alpha,\beta,\gamma,z)=-\frac{1}{2\pi\,i}\frac{\Gamma(1-\alpha)\Gamma(\gamma)}{\Gamma(\gamma-\alpha)}\oint\,(-t)^{\alpha-1}(1-t)^{\gamma-\alpha-1}(1-tz)^{-\beta}dt.
\end{array}
\end{equation}

In order to find a light absorption rates necessarily to solve the integral

\begin{equation}
\begin{array}{c}
J=\int_{-1}^{1}dx\,F(-l,l+1,1,\frac{1-x}{2}).
\end{array}
\end{equation}

Substituting (89) into (90) we obtain the integral in the form

\begin{equation}
\begin{array}{c}
J=-\frac{1}{2\pi\,i}\frac{\Gamma(1+l)\Gamma(1)}{\Gamma(1+l)}\int_{-1}^{1}dx\oint\,(-t)^{-l-1}(1-t)^{l}(1-\frac{t}{2}+\frac{tx}{2})^{-l-1}dt,
\end{array}
\end{equation}

which can be rewritten as follows

\begin{equation}
\begin{array}{c}
J=-\frac{1}{2\pi\,i}\frac{\Gamma(1+l)\Gamma(1)}{\Gamma(1+l)}\int_{-1}^{1}dx\oint\,(-t)^{-l-1}(1-t)^{l}(1-\frac{t}{2})^{-l-1}(1+\frac{tx}{2-t})^{-l-1}dt.
\end{array}
\end{equation}

The solution the following integral

\begin{equation}
\begin{array}{c}
J=\int_{-1}^{1}(1+\frac{tx}{2-t})^{-l-1}dx,
\end{array}
\end{equation}

may be found by substitution

\begin{equation}
\begin{array}{c}
y=\frac{tx}{2-t}.
\end{array}
\end{equation}

We find the solution of the integral

\begin{equation}
\begin{array}{c}
J=\int_{-\frac{t}{2-t}}^{\frac{t}{2-t}}(1+y)^{-l-1}dy=-\frac{2^{-l}}{l+2}(2-t)^{l}+\frac{2^{-l}}{l+2}(1-t)^{-l}(2-t)^{l}.
\end{array}
\end{equation}

Then substituting (95) into (92) we obtain the integral in the form

\begin{equation}
\begin{array}{c}
J=-\frac{1}{2\pi\,i}\frac{\Gamma(1+l)\Gamma(1)}{\Gamma(1+l)}\frac{2^{-l}}{l+2}(\oint\,(-t)^{-l-1}(1-t)^{l}(1-\frac{t}{2})^{-1}dt-\oint\,(-t)^{-l-1}(1-t)^{0}(1-\frac{t}{2})^{-1}dt)2^{l},
\end{array}
\end{equation}

which can be expressed via a hypergeometric functions as follows

\begin{equation}
\begin{array}{c}
J=\frac{\Gamma(1)}{l+2}(F(-l,1,1,\frac{1}{2})\frac{1}{\Gamma(1)}-F(-l,1,-l,\frac{1}{2})\frac{\Gamma(1)}{\Gamma(1+l)\Gamma(-l)}).
\end{array}
\end{equation}

In a similar form can be calculated the integral

\begin{equation}
\begin{array}{c}
J=\int_{-1}^{1}dx\,P_{l}^{m}(x).
\end{array}
\end{equation}

Substituting (88) into (98) we obtain the integral in the form

\begin{equation}
\begin{array}{c}
J=\int_{-1}^{1}dx\,\frac{1}{2^{m}}\frac{(l+m)!}{(l-m)!m!}(1-x^{2})^{m/2}F(m-l,m+l+1,m+1,\frac{1-x}{2}).\\
\end{array}
\end{equation}

Using the formula (89) the integral (99) one can transform into the integral

\begin{equation}
\begin{array}{c}
J=-\frac{1}{2\pi\,i}\frac{\Gamma(1-m+l)\Gamma(m+1)}{\Gamma(1+l)}\frac{1}{2^{m}}\frac{(l+m)!}{(l-m)!m!}\int_{-1}^{1}dx(1-x^{2})^{m/2}\oint\,(-t)^{m-l-1}(1-t)^{l}(1-t\frac{1-x}{2})^{-m-l-1}dt,\\
\end{array}
\end{equation}

which can be rewritten in the form

\begin{equation}
\begin{array}{c}
J=-\frac{1}{2\pi\,i}\frac{\Gamma(1-m+l)\Gamma(m+1)}{\Gamma(1+l)}\frac{1}{2^{m}}\frac{(l+m)!}{(l-m)!m!}\int_{-1}^{1}dx(1-x^{2})^{m/2}\oint\,(-t)^{m-l-1}(1-t)^{l}(1-\frac{t}{2})^{-m-l-1}(1+\frac{tx}{2-t})^{-m-l-1}dt.\\
\end{array}
\end{equation}

In order to find the solution of the integral (101) it is necessarily to consider the integral of form:

\begin{equation}
\begin{array}{c}
J=\int_{-1}^{1}(1-x^{2})^{m/2}(1+\frac{tx}{2-t})^{-m-l-1}dx,
\end{array}
\end{equation}

which can be transformed into the integral of form:

\begin{equation}
\begin{array}{c}
J=(\frac{t}{2-t})^{-m-l-1}\int_{-1}^{1}(1-x^{2})^{m/2}(\frac{2-t}{t}+x)^{-m-l-1}dx.
\end{array}
\end{equation}

The solution the integral (103) one can find using the binomial theorem and following replacements

\begin{equation}
\begin{array}{c}
J=(\frac{t}{2-t})^{-m-l-1}\int_{-1}^{1}\sum_{k=0}^{\gamma}\frac{\gamma!}{k!(\gamma-k)!}(1)^{\gamma-k}(-1)^{k}x^{2k}(\frac{2-t}{t}+x)^{-m-l-1}dx,
\end{array}
\end{equation}

$\gamma=m/2$,

\begin{equation}
\begin{array}{c}
u=(\frac{\frac{2-t}{t}+x}{x})^{-1/(m+l+1)}.
\end{array}
\end{equation}

So integral (104) may be rewritten as follows

\begin{equation}
\begin{array}{c}
J=(\frac{2-t}{t})^{2k-m-l-1}\int_{}^{}u^{2(m+l+1)(k+1)-(m+l+1)^{2}-(m+l+1)}(1-u^{(m+l+1)})^{-2(k+1)+m+l+1}du.
\end{array}
\end{equation}

The solution of the integral (106) one can find by replacement

\begin{equation}
\begin{array}{c}
u=(y)^{1/(m+l+1)}.
\end{array}
\end{equation}

We obtain the following expression for the looking for integral:

\begin{equation}
\begin{array}{c}
J=\frac{1}{(m+l+1)}(\frac{2-t}{t})^{2k-m-l-1}\int_{-t/(2-2t)}^{t/2}y^{2(k+1)-m-l-3+\frac{1}{m+l+1}}(1-y)^{-2(k+1)+m+l+1}dy=\\
=\frac{1}{(m+l+1)}(\frac{2-t}{t})^{2k-m-l-1}\int_{-t/(2-2t)}^{t/2}y^{2(k+1)-m-l-3+\frac{1}{m+l+1}}\times\\
\times\,\sum_{n=0}^{-2k+m+l-1}\frac{(-2k+m+l-1)!}{(n)!(-2k+m+l-1-n)!}(1)^{(-2k+m+l-1-n)}(-y)^{n}dy.\\
\end{array}
\end{equation}

The solution the integral (108) one can find using the binomial theorem

\begin{equation}
\begin{array}{c}
(1-y)^{-2(k+1)+m+l+1}=\sum_{n=0}^{-2k+m+l-1}\frac{(-2k+m+l-1)!}{(n)!(-2k+m+l-1-n)!}(1)^{(-2k+m+l-1-n)}(-y)^{n}.\\
\end{array}
\end{equation}

Substituting equation (109) in the integral (108) one can obtain the looking for integral in the form

\begin{equation}
\begin{array}{c}
J=\frac{1}{(m+l+1)}(\frac{2-t}{t})^{2k-m-l-1}\sum_{n=0}^{-2k+m+l-1}\frac{(-2k+m+l-1)!}{(n)!(-2k+m+l-1-n)!}(1)^{(-2k+m+l-1-n)}(-1)^{n}\times\\
\times\int_{-t/(2-2t)}^{t/2}y^{2(k+1)-m-l-3+\frac{1}{m+l+1}+n}dy.\\
\end{array}
\end{equation}

We find the solution of the integral (110) in the form:

\begin{equation}
\begin{array}{c}
J=\frac{1}{(m+l+1)}(\frac{2-t}{t})^{2k-m-l-1}\sum_{n=0}^{-2k+m+l-1}\frac{(-2k+m+l-1)!}{(n)!(-2k+m+l-1-n)!}(1)^{(-2k+m+l-1-n)}(-1)^{n}\times\\
\times\,\frac{y^{2k-m-l+\frac{1}{m+l+1}+n}}{(2k-m-l+\frac{1}{m+l+1}+n)}|_{-t/(2-2t)}^{t/2}.\\
\end{array}
\end{equation}

Substituting (111) in (104) one can rewrite the integral (104) in the form:

\begin{equation}
\begin{array}{c}
J=(\frac{t}{2-t})^{-m-l-1}\sum_{k=0}^{\gamma}\frac{\gamma!}{k!(\gamma-k)!}(1)^{\gamma-k}(-1)^{k}\times\\
\times\,\frac{1}{(m+l+1)}(\frac{2-t}{t})^{2k-m-l-1}\sum_{n=0}^{-2k+m+l-1}\frac{(-2k+m+l-1)!}{(n)!(-2k+m+l-1-n)!}(1)^{(-2k+m+l-1-n)}(-1)^{n}\times\\
\times\,\frac{y^{2k-m-l+\frac{1}{m+l+1}+n}}{(2k-m-l+\frac{1}{m+l+1}+n)}|_{-t/(2-2t)}^{t/2}.\\
\end{array}
\end{equation}

Substituting (112) in the looking for integral (101) one can rewrite the integral (101) as follows:

\begin{equation}
\begin{array}{c}
J=-\frac{1}{2\pi\,i}\frac{\Gamma(1-m+l)\Gamma(m+1)}{\Gamma(1+l)}\frac{1}{2^{m}}\frac{(l+m)!}{(l-m)!m!}\frac{1}{(m+l+1)}\sum_{k=0}^{\gamma}\frac{\gamma!}{k!(\gamma-k)!}(1)^{\gamma-k}(-1)^{k}\times\\
\times\sum_{n=0}^{-2k+m+l-1}\frac{(-2k+m+l-1)!}{(n)!(-2k+m+l-1-n)!}(1)^{(-2k+m+l-1-n)}(-1)^{n}\times\\
\times\,\sum_{s=0}^{2k}\frac{(2k)!}{(s)!(2k-s)!}(-1)^{s}\times\\
\times\oint\,(-t)^{m-l-1}(1-t)^{l}(1-\frac{t}{2})^{-m-l-1}\times\\
\times\,(\frac{2}{t})^{(2k-s)}\frac{y^{2k-m-l+\frac{1}{m+l+1}+n}}{(2k-m-l+\frac{1}{m+l+1}+n)}|_{-t/(2-2t)}^{t/2}dt,\\
\end{array}
\end{equation}

which can be rewritten in the form

\begin{equation}
\begin{array}{c}
J=-\frac{1}{2\pi\,i}\frac{\Gamma(1-m+l)\Gamma(m+1)}{\Gamma(1+l)}\frac{1}{2^{m}}\frac{(l+m)!}{(l-m)!m!}\frac{1}{(m+l+1)}\sum_{k=0}^{\gamma}\frac{\gamma!}{k!(\gamma-k)!}(1)^{\gamma-k}(-1)^{k}\times\\
\times\sum_{n=0}^{-2k+m+l-1}\frac{(-2k+m+l-1)!}{(n)!(-2k+m+l-1-n)!}(1)^{(-2k+m+l-1-n)}(-1)^{n}\times\\
\times\,\sum_{s=0}^{2k}\frac{(2k)!}{(s)!(2k-s)!}(-1)^{s}\times\\
\times\frac{1}{(2k-m-l+\frac{1}{m+l+1}+n)}\oint\,(-t)^{m-l-1}(1-t)^{l}(1-\frac{t}{2})^{-m-l-1}\times\\
\times\,(\frac{2}{t})^{(2k-s)}((t/2)^{2k-m-l+\frac{1}{m+l+1}+n}-(-t/(2-2t))^{2k-m-l+\frac{1}{m+l+1}+n})dt,\\
\end{array}
\end{equation}

or as follows

\begin{equation}
\begin{array}{c}
J=-\frac{1}{2\pi\,i}\frac{\Gamma(1-m+l)\Gamma(m+1)}{\Gamma(1+l)}\frac{1}{2^{m}}\frac{(l+m)!}{(l-m)!m!}\frac{1}{(m+l+1)}\sum_{k=0}^{\gamma}\frac{\gamma!}{k!(\gamma-k)!}(1)^{\gamma-k}(-1)^{k}\times\\
\times\sum_{n=0}^{-2k+m+l-1}\frac{(-2k+m+l-1)!}{(n)!(-2k+m+l-1-n)!}(1)^{(-2k+m+l-1-n)}(-1)^{n}\times\\
\times\,\sum_{s=0}^{2k}\frac{(2k)!}{(s)!(2k-s)!}(-1)^{s}\times\\
\times\frac{2^{m+l-\frac{1}{m+l+1}-n-s}}{(2k-m-l+\frac{1}{m+l+1}+n)}\oint\,(-t)^{m-l-1}(1-t)^{l}(1-\frac{t}{2})^{-m-l-1}\times\\
\times\,(t)^{s-m-l+\frac{1}{m+l+1}+n}(1-(-1)^{2k-m-l+\frac{1}{m+l+1}+n}(1-t)^{-2k+m+l-\frac{1}{m+l+1}-n})dt.\\
\end{array}
\end{equation}

We find the solution of the looking for integral as follows:

\begin{equation}
\begin{array}{c}
J=\frac{\Gamma(1-m+l)\Gamma(m+1)}{\Gamma(1+l)}\frac{1}{2^{m}}\frac{(l+m)!}{(l-m)!m!}\frac{1}{(m+l+1)}\sum_{k=0}^{\gamma}\frac{\gamma!}{k!(\gamma-k)!}(1)^{\gamma-k}(-1)^{k}\times\\
\times\sum_{n=0}^{-2k+m+l-1}\frac{(-2k+m+l-1)!}{(n)!(-2k+m+l-1-n)!}(1)^{(-2k+m+l-1-n)}(-1)^{n}\times\\
\times\,\sum_{s=0}^{2k}\frac{(2k)!}{(s)!(2k-s)!}(-1)^{s}\times\\
\times\frac{2^{m+l-\frac{1}{m+l+1}-n-s}}{(2k-m-l+\frac{1}{m+l+1}+n)}(-1)^{-s+m+l-\frac{1}{m+l+1}-n}\times\\
\times\,(F(s-2l+\frac{1}{m+l+1}+n,m+l+1,s-l+\frac{1}{m+l+1}+n+1,1/2)\frac{\Gamma(l+1)}{\Gamma(1-s+2l-\frac{1}{m+l+1}-n)\Gamma(s-l+\frac{1}{m+l+1}+n+1)}-\\
-(-1)^{2k-m-l+\frac{1}{m+l+1}+n}F(s-2l+\frac{1}{m+l+1}+n,m+l+1,s+1-2k+m,1/2)\frac{\Gamma(1-2k+m+2l-\frac{1}{m+l+1}-n)}{\Gamma(1-s+2l-\frac{1}{m+l+1}-n)\Gamma(s+1-2k+m)}).\\
\end{array}
\end{equation}

\end{widetext}

\end{document}